\documentclass[journal]{IEEEtran}[12pt]
\ifCLASSINFOpdf
\usepackage[pdftex]{graphicx}
\graphicspath{{../pdf/}{../jpeg/}}
\DeclareGraphicsExtensions{.pdf,.jpeg,.png}
\else
\usepackage[dvips]{graphicx}
\graphicspath{{../eps/}}
\DeclareGraphicsExtensions{.eps}
\fi\usepackage{graphicx}

\usepackage{graphics}
\usepackage{epsfig}
\usepackage{epstopdf}
\usepackage{stfloats}
\usepackage[cmex10]{amsmath}
\usepackage{algorithmic}
\usepackage{array}
\usepackage{mdwmath}
\usepackage{mdwtab}
\usepackage{graphicx}
\usepackage{subfigure}
\usepackage{color}
\usepackage{amsfonts,amssymb}
\usepackage{hyperref} %
\usepackage{multirow}
\usepackage{diagbox}
\usepackage{caption}
\usepackage{makecell}
\usepackage{lineno} 
\usepackage{mathrsfs}
\usepackage{float}  
\usepackage{amssymb,mathrsfs,amsmath}

\usepackage{amsfonts,amsthm,array} 
\usepackage[ruled]{algorithm2e}

\begin{document}
	
	%\linenumbers         %\thanks{Manuscript received.}
	
     \title{Multi-UAV Trajectory Design for Fair and Secure Communication}

    \author{Hongjiang~Lei, %~\IEEEmembership{Senior Member,~IEEE,}	
    	Dongyang~Meng, 
    	Haoxiang~Ran,	
    	Ki-Hong~Park, \\%~\IEEEmembership{Senior Member,~IEEE,}   
    	Gaofeng~Pan %~\IEEEmembership{Senior Member,~IEEE,}  
    	and~Mohamed-Slim~Alouini %,~\IEEEmembership{Fellow,~IEEE}
    	
%    	\thanks{This work was supported by the National Natural Science Foundation of China under Grant 62171031 and 61971080. } %(Corresponding author: \emph{Hongjiang~Lei}).}
    	\thanks{Hongjiang~Lei, Dongyang Meng, and Haoxiang~Ran are with the School of Communications and Information Engineering, Chongqing University of Posts and Telecommunications, Chongqing 400065, China (e-mail: leihj@cqupt.edu.cn, cquptmdy@163.com, rahaxi1958809415@163.com).}
    	\thanks{Gaofeng~Pan is with the School of Cyberspace Science and Technology, Beijing Institute of Technology, Beijing 100081, China (e-mail: gaofeng.pan.cn@ieee.org).}
    	\thanks{K.-H. Park and M.-S. Alouini are with CEMSE Division, King Abdullah University of Science and Technology (KAUST), Thuwal 23955-6900, Saudi Arabia (e-mail: kihong.park@kaust.edu.sa, slim.alouini@kaust.edu.sa).}
    }

\maketitle

%%%%%%%%%%%%%%%%%%%%%%%%%%%%%%%%%%%%%%%%%%%%%%%%%%%%%%%%%%%%%%%%%%%
\begin{abstract}

Unmanned aerial vehicles (UAVs) play an essential role in future wireless communication networks due to their high mobility, low cost, and on-demand deployment. In air-to-ground links, UAVs are widely used to enhance the performance of wireless communication systems due to the presence of high-probability line-of-sight (LoS) links. However, the high probability of LoS links also increases the risk of being eavesdropped, posing a significant challenge to the security of wireless communications. 
In this work, the secure communication problem in a multi-UAV-assisted communication system is investigated in a moving airborne eavesdropping scenario. 
To improve the secrecy performance of the considered communication system, aerial eavesdropping capability is suppressed by sending jamming signals from a friendly UAV. 
An optimization problem under flight conditions, fairness, and limited energy consumption constraints of multiple UAVs is formulated to maximize the fair sum secrecy throughput. 
Given the complexity and non-convex nature of the problem, we propose a two-step-based optimization approach. 
The first step employs the $K$-means algorithm to cluster users and associate them with multiple communication UAVs. 
Then, a multi-agent deep deterministic policy gradient based algorithm is introduced to solve this optimization problem. 
The effectiveness of this proposed algorithm is not only theoretically but also rigorously verified by simulation results. 
		
\end{abstract}


%%%%%%%%%%%%%%%%%%%%%%%%%%%%%%%%%%%%%%%%%%%%%%%%%%%%%%%%%%%%%%%%%%%

\section{Introduction}
\label{sec:introduction}

\subsection{Background and Related Works}

%背景
Unmanned aerial vehicles (UAVs) will play an essential role in future wireless communication networks due to their high mobility, low cost, and on-demand deployment \cite{ZengY2016CM, WuQ2021JSC}. 
In air-to-ground (A2G) links, UAVs are widely used to improve the performance of wireless communication systems due to the existence of line of sight (LoS) links with high probability \cite{LiB2019IOT}. 
With their high mobility, UAVs can be utilized in real-time through trajectory design or location optimization to improve the quality of the A2G channel and provide higher-quality communication services. 
Therefore, the UAV trajectory design plays a decisive role in improving the performance of the UAV-aided communication systems \cite{WuQ2019WCL}.

%单个无人机+非PLS的研究工作
So far, various related works have been done on the trajectory design of UAV-assisted communication systems, which take into account various optimization objectives, such as flight time, energy consumption, fairness, etc. 
Aiming at the problem of data acquisition in widely distributed Internet of Things (IoT) devices, Ref. \cite{ZongJ2019ISWCS} studied the UAV trajectory design and resource allocation scheme under two orthogonal multiple access schemes. 
The task completion time was minimized by jointly optimizing trajectory, power, and time allocation or frequency allocation. 
In Ref. \cite{ZengY2019TWC}, the energy consumption model of rotorcraft UAV was first derived, and then the energy minimization problem of rotary-wing UAVs was studied. 
A method based on path discretization and continuous convex approximation was proposed to solve the problem of minimizing UAV energy consumption optimization by jointly designing the UAV's trajectory, user scheduling, and total task completion time. 
Ref. \cite{BejaouiA2020WCL} considered the max-min fairness problem in A2G communication systems under limited energy consumption and specific quality of service requirements and proposed an iterative algorithm based on sequential convex approximation (SCA) and block coordinate descent (BCD) to solve non-convex problems.

%UAV + PLS
Physical layer security is a supplement to traditional encryption technology based on information theory and can improve information security by utilizing the time-varying characteristics of wireless channels. 
To improve the security of information transmission, UAV communication technology was combined with physical layer security technology in many works \cite{WangHM2019WC, WuQ2019WC, WangJ2022CC}. 
The security of the communication system is enhanced by rationally designing the UAV's flight path as close as possible to legitimate users and away from potential eavesdroppers. 
Ref. \cite{GaoY2021CL} studied the security of energy-constrained rotorcraft UAV-aided communication systems which had internal and external eavesdroppers. 
An iterative algorithm based on SCA and BCD was proposed to solve the non-convex problem by optimizing the trajectory, transmission power of UAVs, and user scheduling. 
Ref. \cite{DuoB2021CC} studied an airborne system with active eavesdropping and maximized the uplink and downlink average secrecy rates (ASRs), respectively, by optimizing the trajectory and transmission power of the UAV.

%UAV +PLS + AN
To enhance the security of wireless communication, one commonly used method is to send artificial noise (AN) to disrupt eavesdroppers. 
For example, AN was utilized in Ref. \cite{MamaghaniTM2021TVT} considering a two-hop UAV-aided communication system. 
The ASR was maximized by jointly optimizing the UAV's trajectory, the transmission power, and the power allocation. 
In Ref. \cite{WangY2021TCCN}, the secrecy performance of cognitive radio networks was improved by utilizing the AN sent by the cognitive UAV to suppress eavesdroppers. 
The total ASR of the cognitive radio network (CRN) was maximized by optimizing subcarrier allocation, UAV's trajectory, and transmission power. 
Ref. \cite{LeiH2023IoTUAV} studied a secure aerial IoT system under friendly jamming UAV collaboration, taking into account the uncertainty of the location of eavesdroppers. 
The ASR was maximized by optimizing user scheduling, UAV trajectories, and transmit power. 
In \cite{ZhangR2021TWC}, two security data collection schemes were proposed for dual-UAV, and the propulsion energy consumption for the UAV was considered. 
The ASR was maximized by jointly optimizing user scheduling, transmission power, and the trajectories of UAVs. 
Moreover, to save onboard energy further and extend flight time, secrecy energy efficiency was maximized.

%UAV + DRL
Compared with traditional convex optimization methods, deep reinforcement learning (DRL)-based technologies are considered to be more suitable for addressing complex challenges in wireless communication. 
The non-convex problems are reformulated as a Markov decision process (MDP) and then trained as a neural network (NN) to learn the mapping from environmental parameters to an optimal solution, which provides a new way to solve the optimization problem of the UAVS communication system.  
Ref. \cite{ZhouX2022WCL} studied minimizing the task completion time in the UAV-assisted data collection system by optimizing the trajectory of UAVs under the constraints of quality of service (QoS) and energy consumption. 
A soft-constrained actor-critic algorithm was proposed by introducing Lagrange's primordial-duality optimization method into the actor-critic framework. 
Ref. \cite{WangY2022IOT} considered UAV-assisted data collection systems in urban environments.  An optimization problem was formulated to minimize the task completion time by optimizing the UAV trajectory. The double delay depth deterministic strategy gradient algorithm was utilized to solve the formulated problem.
Ref. \cite{ZhangM2021WCL} optimized the 3D position and power allocation of the aerial BS to maximize the uplink throughput of the system. 
An algorithm based on deep deterministic policy gradient (DDPG) was developed and combined with water-filling to learn the
optimal 3D position and to obtain optimal power allocation. 
The authors in \cite{YinS2019TVT} studied the communication problem in the case of unknown user location information, transmission power, and channel parameters and proposed a DDPG-based algorithm to maximize the uplink sum rate by optimizing the trajectory of the UAV. 
Ref. \cite{NguyenKK2022TCOM} studied a UAV-assisted IoT system in which the ground user was in a state of random movement. 
A scheme based on the dueling deep Q-network algorithm was proposed to jointly optimize the 3D trajectory and data collection performance of UAVs.

%单个UAV + DRL + 公平性
In aerial communication systems, fairness among users is also a crucial problem that should be considered. 
In \cite{DingR2020TWC}, the authors proposed a novel dynamic band allocation strategy that can consider the flight energy consumption of the UAV and the fairness in the scenario where the user moved randomly. 
They formulated an optimization problem to maximize the uplink fair throughput by optimizing the 3D trajectory and frequency band allocation and utilized the  DDPG algorithm to solve it. 
Ref. \cite{ZhangZ2023IOT} investigated a UAV-assisted communication system with non-orthogonal multiple access (NOMA) technology by designing the 3D trajectory design of UAVs and time allocation in data collection. 
Considering the energy constraints, service quality requirements, and flight conditions, the total fair throughput was maximized, and a  DDPG-based algorithm was proposed to solve the formulated problem.
%单个UAV  +  DRL + PLS
The authors in \cite{DengD2021DCN} jointly optimized the UAV trajectory and transmission power to maximize the total secrecy rate (SR) of the aerial NOMA IoT systems, ensuring the QoS requirements for legitimate users. 
An algorithm based on Q-learning was proposed and combined with SCA technology to solve the optimization problem. 
In \cite{YangH2023TWC}, the achievable SR of UAV systems aided by reflective intelligent surfaces (RIS) was maximized by jointly optimizing the transmit beamforming, AN power, UAV-RIS placement, and RIS' beamforming. 
A post-decision state DQN-based algorithm was proposed to solve the non-convex optimization problem. 

%多个UAV  +  DRL 
Recently, multiple UAVs have been utilized simultaneously to provide high-quality, security, and fair communication services for a wide range of ground users through cooperative effort, and multi-agent reinforcement learning algorithms have become a hot research topic. 
In \cite{XuS2022TVT}, multiple UAVs were utilized for data acquisition tasks in IoT systems. 
Considering the maximum speed and acceleration, collision avoidance, and interference among UAVs, the task completion time was minimized by optimizing the trajectory of multiple UAVs. 
An algorithm based on $K$-means was proposed to assign the task, then a distributed multi-agent DRL algorithm was introduced for the preliminary design of UAV trajectory, and a centralized multi-UAV joint trajectory design DRL algorithm was used to solve the formulated optimization problem. 
In contrast, Ref. \cite{KhodaparastSS2021OJVT} considered UAV flight energy consumption, sensor node energy consumption, and air obstacles constraints and an optimization problem was formulated to minimize the total energy consumption of the considered IoT systems.
The optimization problem was decomposed into three sub-problems: UAV trajectory design, power control during data acquisition, and task assignment among the UAVs. The DDPG and multi-agent DQN algorithms were utilized to realize the UAV trajectory design, power control problems, and the multi-UAV task allocation problem, respectively. 
Ref. \cite{YinS2022IOT} studied a multi-UAV-assisted downlink cellular network in which multiple UAVs served as air base stations (BSs) and provided services to ground users based on frequency division multiple access (FDMA) technology. 
The overall throughput and fair throughput of the system were maximized by optimizing the user scheduling, power, and trajectory of UAVs, respectively. The formulated problem was solved by a multi-agent reinforcement learning algorithm based on DQN. 
Ref. \cite{ZhongR2022TWC} studied the NOMA systems assisted by multiple UAVs to provide services for ground users. 
The total throughput of the system was maximized by jointly optimizing the 3D trajectory design and power distribution of multiple UAVs. 
For the continuously roamed ground users, a $K$-means-based algorithm was utilized to cluster all the users periodically, and a multi-agent reinforcement learning algorithm based on DQN was proposed to obtain the optimal 3D trajectory and power allocation. 
The authors in \cite{DingJR2022IOT} studied the A2G cooperative communication system and focused on jointly optimizing the trajectory and user scheduling of multiple aerial BSs so as to maximize the fair throughput, thereby improving the total throughput and maintaining fairness among users. 
A multi-agent reinforcement learning algorithm based on DDPG was proposed to solve the formulated optimization problem.

\subsection{Motivation and Contributions}

Inspired by the above work, we investigate the joint secure and fairness in a UAV-assisted communication system in a moving airborne eavesdropping scenario.
The aerial eavesdropper is suppressed by a friendly UAV and multiple UAVs are utilized to provide secure communication services to ground users.
Considering the limited energy consumption of the UAVs and the fairness requirements among users, an optimization problem is formulated under the constraints of flight conditions, fairness requirements, and the limited energy consumption of the UAVs.
A two-step optimization method is proposed to maximize the total fair secrecy throughput (FST) by jointly optimizing the trajectories and transmission power of UAVs.
The contributions of this work are summarized as follows.

\begin{enumerate}
	
	\item Secure communication in a multi-UAV-assisted communications system is considered in the presence of a mobile airborne eavesdropper. A jamming UAV is used to suppress the eavesdropping capability of an airborne eavesdropper, thereby assisting multiple communication UAVs to provide secure communication services to ground users.
	A joint multi-UAV trajectory design and power control optimization problem is formulated to maximize the total fair throughput of the network while ensuring energy constraints, fairness and flight conditions.
	
	\item A fair scheduling scheme under reinforcement learning algorithms is considered to obtain a trade-off between maximizing the throughput of the system and the fairness of communication between users. An adaptive energy consumption threshold is utilized to trade-off between the return endpoint and the data collection problem to maximize the total secrecy throughput.
	
	\item A two-step approach is proposed, where the $K$-means algorithm is first used to cluster users and associate them with multiple communication UAVs. Then, a Multi-Agent DDPG (MADDPG) based algorithm is proposed to solve this optimization problem. Our work addresses the dimension imbalance problem, a significant challenge arising from the high role dimensions in the framework. To overcome this, we have designed a dimension expansion mechanism.
	
\end{enumerate}

The rest of the paper is organized as follows.
Section \ref{sec:SystemModel} presents the system model and formulates the corresponding optimization problem.
Section \ref{sec:SCTPD} provides a detailed description of the secure communication through trajectory and power design (SCTPD) algorithm and defines terms related to deep reinforcement learning.
Simulation results and analysis are presented in Section \ref{sec:Simulation}.
Finally, the paper is summarized in Section \ref{sec:Conclusions}.

\section{System Model and Problem Formulation}
\label{sec:SystemModel}

\subsection{System Model}

\begin{figure}[t]
	\centering		
	\includegraphics[width = 2in]{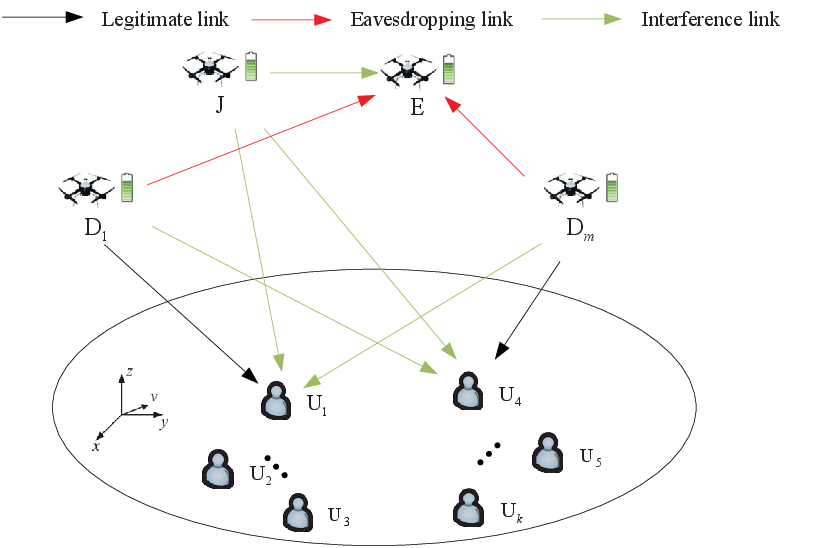}
	\caption{System Model.}
	\label{fig:01}
\end{figure}

As shown in Fig. \ref{fig:01}, we consider a UAV-assisted IoT system in which $M$ UAVs $\left( {{{\rm{D}}_m},m = 1, \cdots, M} \right)$ act as aerial BSs to provide services to $K$ terrestrial users $\left( {{{\rm{U}}_k},k = 1, \cdots, K} \right)$ with a single antenna. 
All the users are divided into $M$ different clusters $\left( {{{\bf{C}}_m},m = 1, \cdots, M} \right)$ in which each cluster is served by one UAV, like \cite{ZhongR2022TWC}. 
There exists a UAV (E) moving from a given start position $\left( {\bf{q}}_{\rm{E}}^{\rm{I}} \right)$ to the end position $\left( {\bf{q}}_{\rm{E}}^{\rm{F}} \right)$ in a straight line during its flight, which is viewed as a potential eavesdropper  \cite{LuW2022TII}.
To improve the security of wireless communication, another UAV $\left( \mathrm{J} \right)$ acts as a friendly jammer to suppress the eavesdropping capability of E through artificial noise \cite{ZhangR2021TWC}.
It is assumed that all UAVs work at a fixed altitude and share the same frequency band, that each UAV serves only one user at a time, and that all terrestrial users are equipped with a single antenna.

Without loss of generality, the flight cycle $T$ is divided equally into $N$ time slots, where each time slot is of length ${\delta _t} = {T}/N$ and $N$ is large enough to ensure that the position of the UAVs can be regarded as approximately constant within each time slot \cite{WuQ2019WC}. 
Furthermore, a Cartesian coordinate system is used to describe the positions of these nodes.
Then, the horizontal position of ${{\rm{D}}_m}$, $\mathrm{E}$, and $\mathrm{J}$ in the $n$th time slot is denoted as ${{\bf{Q}}_{{m}}}\left[ n \right] = {\left[ {{x_{{m}}}\left[ n \right],{y_{{m}}}\left[ n \right]} \right]^T} \in {\mathbb{R}^{2 \times 1}}$,  
${{\bf{Q}}_{\rm{E}}}\left[ n \right] = {\left[ {{x_{\rm{E}}}\left[ n \right],{y_{\rm{E}}}\left[ n \right]} \right]^T} \in {\mathbb{R}^{2 \times 1}}$, 
and 
${{\bf{Q}}_{\rm{J}}}\left[ n \right] = {\left[ {{x_{\rm{J}}}\left[ n \right],{y_{\rm{J}}}\left[ n \right]} \right]^T} \in {\mathbb{R}^{2 \times 1}}$, 
respectively.
The position of ${\mathrm{U}_k}$ is denoted as ${{\bf{Q}}_{\mathrm{U}_k}} \in {\mathbb{R}^{2 \times 1}}$.

The velocities and positions of ${{{\rm{D}}_m}}$ and $\mathrm{J}$ at the $\left(n + 1\right)$th time slot are expressed as \cite{DingR2020TWC}
\begin{align}
	&{v_i}[n + 1] = {v_i}\left[ n \right] + {{\tilde a}_i}\left[ n \right]{\delta _t},\forall i \in \{ \mathbb{M},{\rm{J}}\}, \label{eqn:1}\\
	&{{\bf{Q}}_i}[n + 1] = {{\bf{Q}}_i}\left[ n \right]  \nonumber\\
	&+ \left( {{v_i}\left[ n \right]{\delta _t} + \frac{1}{2}{{\tilde a}_i}\left[ n \right]\delta _t^2} \right){{\bf{v}}_i}\left[ n \right],\forall i \in \{ \mathbb{M},{\rm{J}}\}, \label{eqn:2}
\end{align}
respectively, 
where 
 $\mathbb{M}  \buildrel \Delta \over = \left\{ {1, \cdots ,M} \right\}$, 
${{\tilde a}_i}\left[ n \right] = {\xi _{{a_i}}}\left\| {{{{\bf{\tilde a}}}_i}\left[ n \right]} \right\||$, 
${\xi _{{a_i}}}$ is a binary indicator, 
${\xi _{{a_i}}} =  1$ and ${\xi _{{a_i}}} = -1$ denote that the direction of UAV acceleration is the same as or opposite to the direction of velocity, respectively, 
and 
${{{{\bf{\tilde a}}}_i}\left[ n \right]}$ and ${{\mathbf{v}}_i}\left[ n \right]$ denote the acceleration vector and unit velocity vector of the UAV, respectively.

To avoid possible collisions, the position constraint among UAVs during flight is denoted as
\begin{align}
	\left\| {{{\bf{Q}}_i}\left[ n \right] - {{\bf{Q}}_j}\left[ n \right]} \right\| \ge {\rm{D}},\forall i,j \in \{ \mathbb{M},{\rm{J}},{\rm{E}}\} ,i \ne j,
	\label{eqn:3}
\end{align}
where 
${\rm{D}}$ denotes the minimum distance.

\subsection{Channel Model}

Like \cite{WangY2022IOT, ZhangM2021WCL, DingR2020TWC}, it is assumed the A2G link experiences the probabilistic LoS channel model and the LoS probability in the $n$th slot is expressed as \cite{HouraniA2014WCL}
\begin{align}
	p_{m,{u_k}}^{{\rm{LoS}}}\left[ n \right] = \frac{1}{{1 + {\eta _a}\exp \left( { - {\eta _b}\left( {\arcsin \left( {\frac{H}{{{d_{m,{u_k}}}\left[ n \right]}}} \right) - {\eta _a}} \right)} \right)}},
	\label{eqn:4}
\end{align}
where 
${d_{m,{u_k}}}\left[ n \right] = \sqrt {{{{H}}^2} + {{\left\| {{{\bf{Q}}_m}\left[ n \right] - {{\bf{Q}}_{{{{\rm{U}}_k}}}}\left[ n  \right]} \right\|}^2}}$ represents the distance between ${{{\rm{D}}_m}}$ and  ${{\rm{U}}_k}$ and 
${\eta _a}$ and ${\eta _b}$ are constants related to the propagation environment.
The average path loss between ${{{\rm{D}}_m}}$ and ${{\rm{U}}_k}$ is obtained as 
\begin{align}
		&{L_{m,{u_k}}}\left[ n \right] = p_{m,{u_k}}^{{\rm{LoS}}}\left[ n \right] \times L_{m,{u_k}}^{{\rm{LoS}}}\left[ n \right] + p_{m,{u_k}}^{{\rm{NLoS}}}\left[ n \right] \times L_{m,{u_k}}^{{\rm{NLoS}}}\left[ n \right] \nonumber\\
		&= L_{m,{u_k}}^{{\rm{FS}}}\left[ n \right] + \left( {{\eta _{{\rm{LoS}}}} - {\eta _{{\rm{NLoS}}}}} \right)p_{m,{u_k}}^{{\rm{LoS}}}\left[ n \right] + {\eta _{{\rm{NLoS}}}},
	\label{eqn:4.5}
\end{align}
where 
$p_ {m, {u_k}} ^ {{\rm {NLoS}}} \left[ n \right] = 1 - p_ {m, {u_k}} ^ {{\rm {LoS}}} \left[ n \right] $,  
$L_{m,{u_k}}^{{\rm{FS}}}\left[ n \right] = 20\log {d_{m,{u_k}}}\left[ n \right] + 20\log {f_c} + 20\log \left( {\frac{{4\pi }}{{{v_c}}}} \right)$, 
${f_c} $ as the carrier frequency, 
${v_c} $ for the speed of light, 
${\eta _{{\rm{LoS}}}}$
and 
${\eta _{{\rm{NLoS}}}}$ denote the  path loss of  LoS and NLoS links \cite{HouraniA2014WCL}.
Then, the channel gain between ${{{\rm{D}}_m}}$ and ${{\rm{U}}_k}$ is expressed as
\begin{align}
	{g_{m,{u_k}}}\left[ n \right] = {10^{ - \frac{1}{{10}} \times {L_{m,{u_k}}}\left[ n \right]}}.
	\label{eqn:4.6}
\end{align}
Similarly, 
the channel gain between ${\rm{J}}$ and ${{\rm{U}}_k}$ in the $n$th slot is expressed as
\begin{align}
	{g_{{\rm{J}},{u_k}}}\left[ n \right] = {10^{ - \frac{1}{{10}} \times {L_{{\rm{J}},{u_k}}}\left[ n \right]}}.
	\label{eqn:4.61}
\end{align}
Therefore, the reachable rate between ${{{\rm{D}}_m}}$ and ${{\rm{U}}_k}$ is expressed as 
(\ref{eqn:4.9}), shown at the top of this page, 
where 
${N_0}$ is the noise power spectral density,
$B$ indicates the bandwidth occupied by the user.
${\sum\limits_{i = 1,i \ne m}^M {{p_i}\left[ n \right]{g_{i,{u_k}}}\left[ n \right]} }$ 
represents the interference from other UAVs, 
and 
${{p_{\rm{J}}}\left[ n \right]{g_{{\rm{J}},{u_k}}}\left[ n \right]}$ represents the interference from ${\rm{J}}$.

\begin{figure*}[ht]
	\begin{align}
		{R_{m,{u_k}}}\left[ n \right] = B{\log _2}\left( {1 + \frac{{{p_m}\left[ n \right]{g_{m,{u_k}}}\left[ n \right]}}{{{N_0}B + \sum\limits_{i = 1,i \ne m}^M {{p_i}\left[ n \right]{g_{i,{u_k}}}\left[ n \right] + {p_{\rm{J}}}\left[ n \right]{g_{{\rm{J}},{u_k}}}\left[ n \right]} }}} \right)
		\label{eqn:4.9}
	\end{align}
	\hrulefill
\end{figure*}

The A2A channel gain are expressed as
\begin{subequations}
\begin{align}
	{g_{m,{\rm{E}}}}\left[ n \right] &= \frac{{{\beta _0}}}{{{{\left\| {{{\bf{Q}}_{\rm{E}}}\left[ n \right] - {{\bf{Q}}_m}\left[ n \right]} \right\|}^2}}}\\
	{g_{{\rm{J}},{\rm{E}}}}\left[ n \right] &= \frac{{{\beta _0}}}{{{{\left\| {{{\bf{Q}}_{\rm{E}}}\left[ n \right] - {{\bf{Q}}_{\rm{J}}}\left[ n \right]} \right\|}^2}}}
\end{align}
\end{subequations}
where ${{\beta _0}}$ represents the channel gain at 1m reference distance.
Then, the achievable rate of $E$ is expressed as
(\ref{eqn:4.10}), shown at the top of the next page. 
\begin{figure*}[ht]
	\begin{align}
		{R_{m,{\rm{E}}}}\left[ n \right] = B{\log _2}\left( {1 + \frac{{{p_m}\left[ n \right]{g_{m,{\rm{E}}}}\left[ n \right]}}{{{N_0}B + \sum\limits_{i = 1,i \ne m}^M {{p_i}\left[ n \right]{g_{i,{\rm{E}}}}\left[ n \right]}  + {p_{\rm{J}}}\left[ n \right]{g_{{\rm{J}},{\rm{E}}}}\left[ n \right]}}} \right)
		\label{eqn:4.10}
	\end{align}
	\hrulefill
\end{figure*}

Therefore, the secrecy rate of ${{\rm{U}}_k}$ is expressed as 
\begin{align}
	R_{m,{u_k}}^{\sec }\left[ n \right] = {\left[ {{R_{m,{u_k}}}\left[ n \right] - {R_{m,{\rm{E}}}}\left[ n \right]} \right]^ + },
	\label{eqn:4.11}
\end{align}
where ${\left[ x \right]^ + } \buildrel \Delta \over = \max \left( {x,0} \right)$.
The cumulative secrecy throughput (ST) of a user ${{\rm{U}}_k}$  in cluster ${{\rm{C}}_m}$ at the $n$th time slot is expressed as
\begin{align}
	R_{m,{u_k}}^{{\rm{cum}}}\left[ n \right] = \sum\limits_{j = 1}^n {{b_{m,{u_k}}}\left[ j \right]R_{m,{u_k}}^{\sec }\left[ j \right]{\delta _t}},
	\label{eqn:4.12}
\end{align}
where ${b_{m,{u_k}}}\left[ n \right]$ denotes a binary indicator variable.
Specifically, ${b_{m,{u_k}}}\left[ n \right] = 1$ denotes ${{{\rm{D}}_m}}$ chooses to communicate with ${{\rm{U}}_k}$ where $\sum\nolimits_{{u_i} \in {{\rm{C}}_m}} {{b_{m,{u_i}}}\left[ n \right] \le 1}$.
The cumulative secrecy throughput of all users in cluster ${{\rm{C}}_m}$ at the $n$th time slot is expressed as
\begin{align}
	R_m^{{\rm{cum}}}\left[ n \right] = \sum\limits_{j = 1}^n {\sum\limits_{{u_i} \in {{\rm{C}}_m}} {{b_{m,{u_i}}}\left[ j \right]R_{m,{u_i}}^{\sec }\left[ j \right]{\delta _t}} }.
	\label{eqn:4.13}
\end{align}

\subsection{Energy Model}

UAV energy consumption consists of communication energy consumption and propulsion energy consumption, where the communication energy consumption is usually two orders of magnitude smaller than the flight energy consumption \cite{ZengY2019TWC}. 
Then, the communication energy consumption is ignored, which is a typical operation \cite{DingR2020TWC}.
The propulsion power is expressed as \cite{ZengY2019TWC}
\begin{align}
		{P_i}\left( {{v_i}} \right) &= \underbrace {{P_B}\left( {1 + \frac{{3{v_i}^2}}{{U_{{\rm{tip}}}^2}}} \right)}_{{\rm{bladeprofle}}} \nonumber\\
		&+ \underbrace {{P_I}{{\left( {\sqrt {1 + \frac{{{v_i}^4}}{{4v_0^4}}}  - \frac{{{v_i}^2}}{{2v_0^2}}} \right)}^{1/2}}}_{{\rm{induced}}} \nonumber\\
		&+ \underbrace {\frac{1}{2}{d_0}\rho sA{v_i}^3}_{{\rm{parasite}}},\forall i \in \{ \mathbb{M},{\rm{J}}\}, 
		\label{eqn:4.14}
\end{align}
where
${P_{B}}$ and ${P_{I}}$ are constant parameters indicating the blade profile power and induced power in the hovering state,
$U_{{\mathrm{tip }}}^{}$ is the tip speed of the rotor blades,
${v_0}$ is the average rotor induced speed at hover,
${d_0}$, $s$, $\rho$, and $A$ denote the body drag ratio, rotor stiffness, air density, and rotor disk area, respectively.

The residual energy ${E_i}\left[ n \right]$ of node $i$ ($i \in \{ \mathbb{M},{\rm{J}}\}$) at the $n$th time slot is expressed as
\begin{align}
	{E_i}\left[ n \right] = {E_{{\rm{max}}}} - \sum\limits_{j = 1}^n {{P_i}\left( {{v_i}\left[ j \right]} \right)} {\delta _t},
	\label{eqn:4.15}
\end{align}
where 
${E_{\max }}$  denotes the energy at the initial position, i.e., ${E_i}\left[ 0 \right] = {E_{\max }}$.
To ensure that the UAV can reach the finish line, there must be 
\begin{align}
	{E_i}\left[ N \right] \ge 0,\forall i \in \{ \mathbb{M},{\rm{J}}\}.
	\label{eqn:4.16}
\end{align}

\subsection{Fairness Among Users}

To indicate whether the fairness requirement is satisfied within ${{{\rm{C}}_m}}$ served by ${{\rm{D}}_m}$, 
a binary variable ${I_m}\left[ n \right]$ is defined as \cite{LeiH2024TCOM}
\begin{align}
	{I_m}\left[ n \right] = \left\{ {\begin{array}{*{20}{l}}
			{1,}&{{{\hat f}_m}\left[ {n - 1} \right] \ge {k_f}\;\;{\rm or}\;\;{I_m}[n - 1] = 1,}\\
			{0,}&{{{\hat f}_m}\left[ {n - 1} \right] \le {k_f}\;\;{\rm and}\;\;{I_m}[n - 1] = 0.}
	\end{array}} \right.
	\label{eqn:4.17}
\end{align}
where 
${{\hat f}_m}\left[ {n - 1} \right] = \frac{1}{{\left| {{{\rm{C}}_m}} \right|\sum\limits_{{u_i} \in {{\rm{C}}_m}} {{{\left( {\frac{{R_{m,{u_i}}^{{\rm{cum}}}\left[ {n - 1} \right]}}{{R_m^{{\rm{cum}}}\left[ {n - 1} \right]}}} \right)}^2}} }}$
denotes the Jain fairness index at the $\left( {n - 1} \right)$th slot, 
${k_f}$ is the target fairness threshold, 
${I_m}\left[ n \right] = 1$ 
and 
${I_m}\left[ n \right] = 0$ 
denote that the fairness requirement in ${{\rm{C}}_m}$ is satisfied or not satisfied, respectively.
It must be noted that 
${I_m}\left[ n \right]$ depends on ${{\hat f}_m}\left[ {n - 1} \right]$ and ${I_m}\left[ {n - 1} \right]$.
When ${{\hat f}_m}\left[ {n - 1} \right] \ge {k_f}$ or ${I_m}\left[ {n - 1} \right] = 1$, there is ${I_m}\left[ n \right] = 1$.
These two conditions indicate that the fairness constraint is satisfied at the $\left( {n - 1} \right)$th time slot and that the fairness constraint has been satisfied before the $\left( {n - 1} \right)$th time slot, respectively.

Like \cite{LeiH2024TCOM}, the fairness factor of ${{\rm{U}}_k}$ in cluster ${{\rm{C}}_m}$ at the $n$th time slot is defined as 
(\ref{eqn:4.18}), shown at the top of this page,
where 
${k_{{\rm{gp}}}} > 0$
and 
${R_{\max }}$ denote  the attenuation coefficient and
the secrecy throughput threshold, respectively.
The fairness factor is explained as follows: 
when ${{I_m}\left[ n \right] = 0}$, ${I_{m,{u_k}}}\left[ n \right]$ decreases as the cumulative secrecy throughput $R_{m,{u_k}}^{{\rm{cum}}}\left[ n - 1 \right]$ increases. 
In this case, 
the scheduling scheme of ${{{\rm{D}}_m}}$ is influenced by ${{I_m}\left[ n \right]}$ and tends to prioritize access to users with higher ${I_{m,{u_k}}}\left[ n \right]$, thus going to improve the fairness among users.
\begin{figure*}[ht]
	\begin{align}
		{I_{m,{u_k}}}\left[ n \right] = \left\{ {\begin{array}{*{20}{l}}
			{2{{\left( {1 + {e^{\left( {R_{m,{u_k}}^{{\rm{cum}}}\left[ {n - 1} \right] - {R_{\max }}} \right)*{k_{{\rm{gp}}}}}}} \right)}^{ - 1}} - 1,}&{{I_m}\left[ n \right] = 0}\\
			{1,}&{{I_m}\left[ n \right] = 1}
		\end{array}} \right.
		\label{eqn:4.18}
	\end{align}
	\hrulefill
\end{figure*}

A fair scheduling scheme is proposed in this work. 
Specifically, ${{{\rm{D}}_m}}$ chooses to establish a communication link with the user with the maximum fair secrecy rate (${I_{m,{u_j}}}\left[ n \right]R_{m,{u_j}}^{\sec }\left[ n \right]$) at each time slot, thus the total FST is maximized.
The scheduling index ${b_{m,{u_k}}}\left[ n \right]$ between ${{{\rm{D}}_m}}$ and ${{\rm{U}}_k}$ in the $n$th time slot is defined as
\begin{align}
	{b_{m,{u_k}}}\left[ n \right] = \left\{ {\begin{array}{*{20}{l}}
			{1,}&{{\rm{if}}\;\;{u_k} = \mathop {\arg \max }\limits_{{u_j} \in {{\rm{C}}_m}} {I_{m,{u_j}}}\left[ n \right]R_{m,{u_j}}^{\sec }\left[ n \right],}\\
			{0,}&{{\rm{otherwise}}.}
	\end{array}} \right.
	\label{eqn:4.19}
\end{align}

More specifically, in the $n$th time slot, ${{{\rm{D}}_m}}$ first determines whether the fairness constraint is satisfied or not, based on the user's cumulative secrecy throughput in the $\left( {n - 1} \right)$th time slot.
If the fairness constraint given in (\ref{eqn:4.17}) is not satisfied, ${{{\rm{D}}_m}}$ will schedule unvisited users to improve the fairness among users.
Otherwise, the problem becomes a trade-off problem between fairness and secrecy throughput.

The sum FST of all the UAVs at the $n$th time slot is expressed as
\begin{align}
	{R^{{\rm{ins}}}}\left[ n \right] = \sum\limits_{m = 1}^M {\sum\limits_{{u_i} \in {{\rm{C}}_m}} {{I_{m,{u_i}}}\left[ n \right]{b_{m,{u_i}}}\left[ n \right]R_{m,{u_i}}^{\sec }\left[ n \right]{\delta _t}} }.
	\label{eqn:4.20}
\end{align}
Then the FST of the system over the whole mission period is expressed as
\begin{align}
	{R^{{\rm{sum}}}} = \sum\limits_{n = 1}^N {\sum\limits_{m = 1}^M {\sum\limits_{{u_i} \in {{\rm{C}}_m}} {{I_{m,{u_i}}}\left[ n \right]{b_{m,{u_i}}}\left[ n \right]R_{m,{u_i}}^{\sec }\left[ n \right]{\delta _t}} } }.
	\label{eqn:4.21}
\end{align}

\subsection{Optimization problem}

The goal of this work is to maximize the FST of the considered systems by jointly designing the trajectories and the transmit power of the UAVs.
Let 
${\bf{Q}} = \left\{ {{{\bf{Q}}_1}\left[ n \right], \cdots ,{{\bf{Q}}_M}\left[ n \right],{{\bf{Q}}_{\rm{J}}}\left[ n \right],\forall n} \right\}$,
${\bf{P}} = \left\{ {{p_1}\left[ n \right], \cdots ,{p_M}\left[ n \right],{p_{\rm{J}}}\left[ n \right],\forall n} \right\}$.
%根据无人机数量分蔟，考虑簇内公平性
The optimization problem is formulated as %\footnote{\textbf{\color{red}fairness inter-cluster}}
\begin{subequations}
	\begin{align}	
		{\mathcal{P}_{1}}: &\mathop {\max }\limits_{{\bf{Q}},{\bf{P}}} {R^{{\rm{sum}}}}, \label{p1a} \\  
		{\mathrm{s.t.}}\;
		&{\rm{C1}}: {E_i}\left[ 0 \right] = {E_{\max }},{E_i}\left[ N \right] \ge 0,\forall i \in \{ \mathbb{M},{\rm{J}}\}, \label{p1b}\\
		&{\rm{C2}}: {{\bf{Q}}_i}\left[ 0 \right] = {\bf{q}}_i^{\rm{I}},{{\bf{Q}}_i}\left[ N \right] = {\bf{q}}_i^{\rm{F}},\forall i \in \{ \mathbb{M},{\rm{J}}\}, \label{p1c}\\
		&{\rm{C3}}: \left\| {{{\bf{Q}}_i}\left[ n \right] - {{\bf{Q}}_j}\left[ n \right]} \right\| \ge {\rm{D}}, \notag \\
		& \,\,\,\,\,\,\,\,\,\,\,\, \forall n,\forall i,j \in \{ \mathbb{M},{\rm{J}},{\rm{E}}\} ,i \ne j, \label{p1d}\\
		&{\rm{C4}}: {{\bf{Q}}_i}[n + 1] = {{\bf{Q}}_i}\left[ n \right] \notag \\
		& \,\,\,\, + \left( {{v_i}\left[ n \right]{\delta _t} + \frac{1}{2}{{\tilde a}_i}\left[ n \right]\delta _t^2} \right){{\bf{v}}_i}\left[ n \right],\forall i \in \{ \mathbb{M},{\rm{J}}\},  \label{p1e}\\		
		&{\rm{C5}}: {v_i}\left[ 0 \right] = 0,\forall i \in \{ \mathbb{M},{\rm{J}}\}, \label{p1f}\\
		&{\rm{C6}}: 0 \le {v_i}\left[ n \right] \le {v_{\max }},\forall n,\forall i \in \{ \mathbb{M},{\rm{J}}\}, \label{p1g}\\
		&{\rm{C7}}: {a_{\min }} \le {{\tilde a}_i}\left[ n \right] \le {a_{\max }},\forall n,\forall i \in \{ \mathbb{M},{\rm{J}}\}, \label{p1h}\\
		&{\rm{C8}}: 0 \le {{p_m}\left[ n \right]} \le {P_{\rm{u}}^{\max} },\forall n, \forall m \in \{ \mathbb{M}\}, \label{p1i}\\    																		
		&{\rm{C9}}: 0 \le {{p_{\rm{J}}}\left[ n \right]} \le {P_{\rm{J}}^{\max} },\forall n,   				        			\label{eq:1j}\\
		&{\rm{C10}}: (\ref{eqn:4.17}), (\ref{eqn:4.18}), (\ref{eqn:4.19}), \notag    
	\end{align}
\end{subequations}
where 
$P_{\rm{u}}^{\max}$ and $P_{\rm{J}}^{\max}$ denote the peak transmit power of UAVs,
$\rm{C1}$ denotes the energy consumption constraints, 
$\rm{C2}$ denotes the constraints of the initial and final positions,
$\rm{C3}$ denotes the safe distance constraint between all the UAVs, 
$\rm{C4}$, $\rm{C5}$, and $\rm{C6}$ denote the velocity constraints,
$\rm{C7}$ denotes the acceleration constraints, 
$\rm{C8}$ and $\rm{C9}$ are the peak power constraints of transmitting and jamming signals in each time slot, respectively,
and 
$\rm{C10}$ denotes the fairness constraints and user scheduling constraints.

It is challenging to solve $\mathcal{P}_{1}$ since the optimization objective in $\mathcal{P}_{1}$ is a non-convex multivariate coupled problem with respect to $\mathbf{Q}$ and $\mathbf{P}$. The energy constraint $\rm{C1}$ and the distance constraint $\rm{C4}$ are also nonconvex for $\mathbf{Q}$.
In addition, probabilistic LoS links are considered, where the channel gain depends not only on the trajectory but also on the distance between the UAV and the user, the flight altitude, and the probability of the LoS link.
The factors discussed above contribute to the complexity of the total throughput between the UAVs and users. These complexities challenge traditional convex optimization methods, suggesting the need for alternative approaches to solve this problem effectively. 
Since the optimization problem involves the trajectory optimization of all the UAVs and the optimization variables are continuous.
To solve $\mathcal{P}_{1}$, a multi-agent deep reinforcement learning (MADRL) algorithm is proposed where the powerful learning capability of DRL is utilized to find the optimal solution by exploring a vast policy space.

\section{Secure Communication through Trajectory and Power Design Algorithm}
\label{sec:SCTPD}

This section presents a solution for the multi-UAV trajectory optimization and power control problem in the aerial eavesdropping scenario.
The solution consists of two steps. Firstly, a $K$-means algorithm is utilized to cluster and correlate multiple communication UAVs with ground users.
Then, a MADDPG-based algorithm named SCTPD is proposed, 
which aims to enable multiple communicating UAVs to provide secure and fair communication services by jointly optimizing the trajectory and power of various UAVs.

In the following subsections, {an overview of} the $K$-means algorithm is first given, then the difficulties in MADRL and the reconstruction of the optimization problem using the partially observable Markov decision process (POMDP) are briefly described.
The state space, action space, reward design, and network framework in the SCTPD algorithm are described in detail.
Finally, the specific training process of the SCTPD algorithm is given.

\subsection{$K$-means-based User Clustering Algorithm}

The $K$-means-based user clustering algorithm is heuristic and has a better user clustering performance in wireless communications \cite{XuS2022TVT}.
Since multiple communication UAVs ${{{{\rm{D}}}_m}}$ serve the ground users in an area at the same time, it may cause significant interference to the communication links of the ground users, as well as in order to avoid collisions between multiple UAVs as much as possible.
Therefore, the $K$-means algorithm is utilized to cluster ground users based on their spatial locations and correlate them with multiple UAVs to minimize the risk of interference and collision with ground users.
In this algorithm, the set of user positions ($\left\{ {{{\bf{Q}}_{{{\rm{U}}_1}}},{{\bf{Q}}_{{{\rm{U}}_2}}}, \cdots ,{{\bf{Q}}_{{{\rm{U}}_k}}}} \right\}$) is input as an observation set to each UAV and the user partition is broadcast to all UAVs through the control channel \cite{ZhongR2022TWC}.
The set of user locations is partitioned into $M$ clusters based on the spatial distance between users in order to keep the users within the clusters as closely connected as possible. 
The sum-of-squares error is expressed as
\begin{align}
	{\rm{ESS}} = \sum\limits_{m = 1}^M {\sum\nolimits_{{{\bf{Q}}_{{{\rm{U}}_i}}} \in {{\bf{C}}_m}} {{{\left\| {{{\bf{Q}}_{{{\rm{U}}_i}}} - {{\bf{u}}_m}} \right\|}^2}} } ,
	\label{eqn:4.23}
\end{align}
where 
${{\bf{u}}_m}$ denotes the average vector of clusters ${{\bf{C}}_m}$. 
The $K$-means algorithm is utilized to find the cluster $\left\{ {{{\bf{C}}_1},{{\bf{C}}_2}, \cdots ,{{\bf{C}}_M}} \right\}$ with the smallest ESS. 
The specific clustering process is to first randomly select $M$ locations as the initial cluster centers, then assign each user to the nearest cluster and recalculate the center of mass of each cluster, iterating this process repeatedly until the centers of mass of all the clusters no longer change.
In this work, it assumed that all the UAVs have different initial positions. When the clustering algorithm converges, each cluster is assigned to a single communication UAV to provide secure and fair communication services based on the distance from the UAV to the center of each cluster.

\subsection{Optimization problem based on POMDP reconstruction}

In a multi-agent environment, the state of each agent changes during the training process, and the decisions and behaviors of each agent have an impact on the states and behaviors of other agents.
The DRL algorithms designed for the single agent can not be utilized directly in multi-agent scenarios.  
To address this challenge, the MADDPG-based SCTPD algorithm is proposed in this section to solve for the trajectory and power of multiple UAVs.

In a multi-agent scenario, each agent usually only observes part of the environment information and chooses its actions independently and cannot access to the state of the whole system.
Therefore, in this section, POMDP is utilized to model the system to improve the ability to understand and adapt to the environment.
Each UAV is considered as agent, which is viewed as a POMDP and represented by 
$\left\langle {{\bf{S}},{\bf{A}},{\bf{R}},{\bf{P}},\gamma } \right\rangle$,
where 
${\bf{S}} = \left\{ {{o_1}, \cdot  \cdot  \cdot ,{o_M},{o_{\rm{J}}}} \right\}$ denotes the global state information of the environment, 
${o_m}$ $\left( {\forall m} \right)$ and ${o_{\rm{J}}}$ denote the partially observable state information of the communicating UAV and the jamming UAV, respectively,
${\bf{A}} = {a_1} \times  \cdot  \cdot  \cdot  \times {a_M} \times {a_{\rm{J}}}$ denotes the joint multi-agent action space,  
${a_m}$ $\left( {\forall m} \right)$ and ${a_{\rm{J}}}$ denote the action spaces of the communicating and the jamming UAVs, respectively,
${\bf{R}}$ denotes the reward space, 
${\bf{P}}$ denotes the state transfer probability function, 
and 
$\gamma$ denotes the discount factor.

In the following subsections, we first define the state, action, and reward space. 
Then we introduce the four NNs: the actor networks, the critic networks, and their respective target networks, which are utilized to learn the decision policy. Finally, the training algorithm is given.

\subsection{State Space}

It should be noted that two different roles of UAVs (communication and jamming UAVs)  are involved in the optimization problem of this work. 
The state space of ${{{\rm{D}}_m}}$ consists of 
the positions of $J$ and $E$, 
the positions of all the communication UAVs, the cumulative secrecy throughput of ${{{\rm{D}}_m}}$ with respect to all the users in the cluster ${{\mathbf{C}_m}}$, the Euclidean distances from the endpoint, the speed, and the remaining energy consumption.
Let
$\widehat {\bf{Q}}\left[ n \right] = \left\{ {{{\bf{Q}}_1}\left[ n \right], \cdots ,{{\bf{Q}}_M}\left[ n \right],{{\bf{Q}}_{\rm{J}}}\left[ n \right],{{\bf{Q}}_{\rm{E}}}\left[ n \right]} \right\}$ and 
$\widehat {\bf{R}}_m^{{\rm{cum}}}\left[ n \right] = \left\{ {R_{m,{u_i}}^{{\rm{cum}}}\left[ n \right],\forall {u_i} \in {{\bf{C}}_m}} \right\}$.
Therefore, the partially observed state space of ${{{{\rm{D}}}_m}}$ is expressed as
\begin{align}
	{o_m}\left[ n \right] = \left\{ {\widehat {\bf{Q}}\left[ n \right],\widehat {\bf{R}}_m^{{\rm{cum}}}\left[ n \right],\left\| {{{\bf{Q}}_m}\left[ n \right] - {\bf{q}}_m^{\rm{F}}} \right\|,{v_m}\left[ n \right],{E_m}\left[ n \right]} \right\}.
	\label{eqn:4.24}
\end{align}
One can observe that  ${{{{\rm{D}}}_m}}$ has a total of $\left( {2M + 7 + \left| {{{\bf{C}}_m}} \right|} \right)$ state dimensions.

The state space of ${\rm{J}}$ consists of the current positions of all communicating UAVs, the position of $J$ and $E$, the Euclidean distance from the endpoint, the velocity and the residual energy consumption.
Therefore, the partially observed state space of ${\rm{J}}$ is expressed as
\begin{align}
	{o_{\rm{J}}}\left[ n \right] = \left\{ {\widehat {\bf{Q}}\left[ n \right],\left\| {{{\bf{Q}}_{\rm{J}}}\left[ n \right] - {\bf{q}}_{\rm{J}}^{\rm{F}}} \right\|,{v_{\rm{J}}}\left[ n \right],{E_{\rm{J}}}\left[ n \right]} \right\}.
	\label{eqn:4.25}
\end{align}
One can observed that  $J$ has a total of $\left( {2M + 7} \right)$ state dimensions.

It is worth noting that the state space of each agent includes the positional information of all UAVs.
However, 
each agent's state space contains only its own flight speed, energy consumption, and Euclidean distance from the endpoint.
The state space of $J$ does not include the cumulative secrecy throughput of all users.

\subsection{Action Space}

The action spaces of ${{{\rm{D}}_m}}$ and ${\rm{J}}$ consist of UAV trajectories and transmit power.
Since the trajectory can be obtained by iterating through the velocity vector, the velocity vector of the UAV is used as the action to represent the trajectory equivalently.
Spherical coordinates $\{ v,\varphi \}$ are used to describe the speed and flight direction of the UAV, where $- \pi \le \varphi \le \pi$ denotes the azimuth angle \cite{DingR2020TWC}.
Thus, the action spaces of ${{{\rm{D}}_m}}$ and ${\rm{J}}$ is expressed as 
\begin{align}
	{{{a}}_i}\left[ n \right] = \left\{ {{v_i}\left[ n \right],{\varphi _i}\left[ n \right],{p_i}\left[ n \right]} \right\},\forall i \in \left\{ {\mathbb{M},{\rm{J}}} \right\}
	\label{eqn:4.26}
\end{align}
One can observed that each agent has a total of 3 action dimensions.

\subsection{Reward Design} 

In DRL, the reward is utilized to evaluate state behavior, and the optimization problem is translated into the maximization of cumulative rewards through reward design \cite{NgYA1999LCML}.
In the proposed SCTPD algorithm, the reward function consists of four components: 
the energy consumption reward, 
the reach-endpoint reward, 
the FST reward, 
and 
the constraint reward.

\subsubsection{The energy consumption reward}

The energy consumption reward is utilized to encourage UAVs to control their energy consumption during flight by optimizing their speed. 
Thus, the total FST is maximized with limited energy consumption. 
The energy consumption reward is defined as
\begin{align}
	{r_{i,{{\rm{ec}}}}}\left[ n \right] =  - {\kappa _{{{\rm{ec}}}}}{P_i}\left[ n \right]{\delta _t},\forall i \in \left\{ {\mathbb{M},{\rm{J}}} \right\},
	\label{eqn:4.27}
\end{align}
where
the subscript `${\mathrm{ec}}$' denotes energy consumption and
${\kappa _{{\mathrm{ec}}}}$ is a positive constant that adjusts the size of the reward for the energy consumption component.

\subsubsection{The reach endpoint reward}

To address the sparse rewards during the return to the endpoint, this work utilizes a reward shaping technique, wherein more continuous and informative feedback is provided, and learning efficiency is enhanced, making it easier for UAVs to reach the endpoint.
To characterize the state in which a UAV is about to run out of energy and must return to the endpoint, a binary indicator ${\xi _{i,{{\rm{rd}}}}}\left[ n \right],\forall i \in \left\{ {\mathbb{M},{\rm{J}}} \right\}$ is defined as 
\begin{align}
	{\xi _{i,{{\rm{rd}}}}}\left[ n \right] = \left\{ {\begin{array}{*{20}{l}}
			{1,}&{{\rm{if}}\;\;{E_i}\left[ n \right] \le {E_{i,\min }},}\\
			{0,}&{{\rm{otherwise}},}
	\end{array}} \right.
	\label{eqn:4.28}
\end{align}
where ${E_i}\left[ n \right]$ is given in (\ref{eqn:4.16}) and ${E_{i,\min }}$ is an energy threshold. 
To enable UAVs to maximize the FST, based on the relationship between the Euclidean distance to the endpoint and the energy threshold, an adaptive energy threshold is defined as \cite{LeiH2024TCOM}
\begin{align}
	{E_{i,\min }} = \frac{{P\left( {{v_{{\rm{mr}}}}} \right) \cdot \left\| {{{\bf{Q}}_i}\left[ n \right] - {\bf{q}}_i^{\rm{F}}} \right\|}}{{{v_{{\rm{mr}}}}}} + {E_0}, \forall i \in \left\{ {\mathbb{M},{\rm{J}}} \right\}
	\label{eqn:4.29}
\end{align}
where 
${{v_{{\mathrm{mr}}}}} = \arg \mathop {\min }\limits_{v \ge 0} \frac{{P(v)}}{v}$ denotes the optimal speed for the %maximum distance traveled for a given energy 
minimum consumed energy per unit distance travel 
and 
${E_0}$ is an energy compensation constant.
Therefore, like \cite{LeiH2024TCOM}, the successive reward {of multiple UAVs about the endpoints} is defined as 
(\ref{eqn:4.30}), shown at the top of the next page, 
where 
$\Delta {d_i} = \left\| {{{\bf{Q}}_i}\left[ n \right] - {\bf{q}}_i^{\rm{F}}} \right\| - \left\| {{{\bf{Q}}_i}\left[ {n + 1} \right] - {\bf{q}}_i^{\rm{F}}} \right\|$
indicates the difference between the distance between the UAV and the endpoint after moving one time slot, 
${{\kappa _{{\mathrm{rd1}}}}}$, ${{\kappa _{{\mathrm{{{\rm{rd}}}2}}}}}$, and ${{\kappa _{{\mathrm{rd3}}}}}$ are constants used to adjust the magnitude of the incremental distance rewards, the distance rewards, and the slope of the rewards with distance, respectively. 
In particular, when ${\xi _{i,{{\rm{rd}}}}}\left[ n \right] = 1$, 
the successive reward associated with the endpoint consists of 
${\kappa _{{\rm{rd1}}}}\Delta {d_i}$ 
and 
$\frac{{{\kappa _{{\rm{rd}}2}}}}{{1 + {\kappa _{{\rm{rd}}3}}\left\| {{{\bf{Q}}_i}\left[ n \right] - {\bf{q}}_i^{\rm{F}}} \right\|}}$. 
The former denotes the incremental distance reward of the flight toward the endpoint and the latter denotes the distance reward of the UAV with respect to the endpoint position.

\begin{figure*}[ht]
	\begin{align}
		{r_{i,{\rm{rd}}}}\left[ n \right] = \left( {{\kappa _{{\rm{rd}}1}}\Delta {d_i} + \frac{{{\kappa _{{\rm{rd}}2}}}}{{1 + {\kappa _{{\rm{rd}}3}}\left\| {{{\bf{Q}}_i}\left[ n \right] - {\bf{q}}_i^{\rm{F}}} \right\|}}} \right){\xi _{i,{\rm{rd}}}}\left[ n \right], \forall i \in \left\{ {\mathbb{M},{\rm{J}}} \right\}
		\label{eqn:4.30}
	\end{align}
	\hrulefill
\end{figure*}

To indicate whether the UAV arrives at the endpoint when the energy is exhausted, a discrete arrival reward $\left( {{r_{i,{\rm{ar}}}}\left[ n \right], i \in \left\{ {\mathbb{M},{\text{J}}} \right\}} \right)$ is defined as
\begin{align}
	{r_{i,{\rm{ar}}}}\left[ n \right] = \left\{ {\begin{array}{*{20}{c}}
			{0,}&{n < N,}\\
			{{\xi _{i,{\rm{ar}}}}{\kappa _{{\rm{ar}}}} + \left( {1 - {\xi _{i,{\rm{ar}}}}} \right){\kappa _{{\rm{nar}}}},}&{n = N,}
	\end{array}} \right.
	\label{eqn:4.31}
\end{align}
where 
${\xi _{i,{\rm{ar}}}} = 1$ means that the drone reaches the endpoint in the last time slot, 
otherwise ${\xi _{i,{\rm{ar}}}} = 0$,  
${\kappa _{{\rm{ar}}}}$ is a positive constant that encourages the UAV to reach the endpoint, 
and 
${\kappa _{{\rm{nar}}}}$ is a negative constant that penalizes the UAV which is not reaching.

\subsubsection{The fair secrecy throughput reward} 

${{{\rm{D}}_m}}$ works in the same frequency band such that they interfere with each other, it is necessary for all the UAVs collaborate with each other by optimizing the transmit power to maximize the total FST. 
The goal of the $J$ is to increase the secrecy rate of ${{{\rm{D}}_m}}$ by suppressing eavesdropping. 
Therefore, the rewards of ${{{\rm{D}}_m}}$ and ${\rm{J}}$ is defined as
\begin{align}
		{r_{i,\sec }}\left[ n \right] &= \left( {\left( {1 - {\xi _{i,{{\rm{rd}}}}}\left[ n \right]} \right){\kappa _{{\rm{th}}}} + {\xi _{i,{{\rm{rd}}}}}\left[ n \right]{\kappa _{{\rm{nth}}}}} \right) \nonumber\\
		& \times {R^{{\rm{ins}}}}\left[ n \right],\forall i \in \left\{ {\mathbb{M},{\rm{J}}} \right\},
		\label{eqn:4.32}
\end{align}
where 
${\kappa _{{\rm{th}}}}$ denotes the reward when energy is sufficient 
and 
${\kappa _{{\rm{nth}}}}$ signifies the reward when the flight energy is about to be exhausted and a return to the endpoint is required. 
It is worth noting that ${\kappa _{{\rm{th}}}} > {\kappa _{{\rm{nth}}}}$ because the reward is reduced when the energy is about to be exhausted to make the UAV tend to return to the endpoint.

\subsubsection{Constraint rewards} 

To penalize the UAV for violating the constraints given in (\ref{p1d}) and (\ref{p1h}), the following rewards are defined as
\begin{subequations}
	\begin{align}
		{r_{i,{\rm{A}}}}\left[ n \right] = {\xi _{i,{\rm{A}}}}\left[ n \right]{\kappa _{\rm{A}}},\forall i \in \left\{ {\mathbb{M},{\rm{J}}} \right\}, \label{conAcce}\\
		{r_{i,{\rm{L}}}}\left[ n \right] = {\xi _{i,{\rm{L}}}}\left[ n \right]{\kappa _{\rm{L}}},\forall i \in \left\{ {\mathbb{M},{\rm{J}}} \right\}, \label{conDis}
	\end{align}
\end{subequations}
where 
${\xi _{i,{\rm{A}}}} = 1$ and ${\xi _{i,{\rm{L}}}} = 1$ denote unsatisfied acceleration and distance constraints, respectively. 
Negative constants ${\kappa _{\rm{A}}}$ and ${\kappa _{\rm{L}}}$ indicate specific rewards for violating the acceleration constraints and distance constraints.

To summarize, in the SCTPD algorithm, the overall reward of ${{{\rm{D}}_m}}$ and ${\rm{J}}$ is expressed as 
\begin{align}
		{r_i}\left[ n \right] &= {r_{i,{\rm{ec}}}}\left[ n \right] + {r_{i,{\rm{rd}}}}\left[ n \right] + {r_{i,{\rm{ar}}}}\left[ n \right] + {r_{i,\sec }}\left[ n \right] \nonumber \\
		& \,\,\,\,\ + {r_{i,{\rm{A}}}}\left[ n \right] + {r_{i,{\rm{L}}}}\left[ n \right],\forall i \in \left\{ {\mathbb{M},{\rm{J}}} \right\}.
		\label{eqn:4.35}
\end{align}

\subsection{Network framework}

\begin{figure*}[t]
	\centering
	\includegraphics[width = 5in]{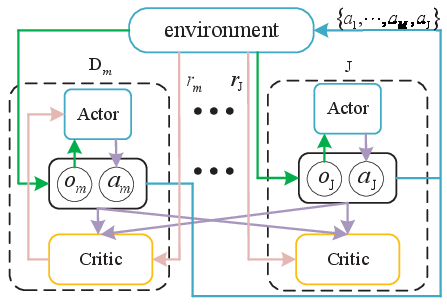}
	\caption{MADDPG architecture.}
	\label{fig02}
\end{figure*}

Fig. \ref{fig02} shows the network architecture of the MADDPG algorithm.
In the SCTPD algorithm, ${{{\rm{D}}_m}}$ and ${\rm{J}}$ are considered as agents and the considered communication system is considered as the environment.
The goal of each agent is to select the best action based on its experience to maximize the cumulative reward over time \cite{XuS2022TVT}. 
Unlike the network architecture of the DDPG algorithm, each agent in MADDPG has its actor network, critic network, and the corresponding target network.
The actor network of each agent utilizes independent observations as inputs to generate actions, while the critic network of each agent uses observations and actions of all agents as inputs.
In this way, the critic network is enabled to evaluate the joint actions of multiple agents more accurately, which helps the agents to make more accurate decisions.

This subsection focuses on the training process for stabilizing the MADDPG by dimensional extension in the SCTPD algorithm and its implementation in actor networks, critic networks, and their respective target networks.

\textbf{\textbf{1. Actor Network}}:
In the SCTPD algorithm, the action of each agent is first normalized as ${v_i}\left[ n \right] = \frac{{\left( {{\lambda _{{v_i}}} + 1} \right){v_{\max }}}}{2},{\varphi _i}\left[ n \right] = \frac{{\left( {{\lambda _{{\varphi _i}}} + 1} \right)\pi }}{2},{P_i}\left[ n \right] = \frac{{\left( {{\lambda _{{p_i}}} + 1} \right){P_{\max }}}}{2},\forall i \in \left\{ {\mathbb{M},{\rm{J}}} \right\}$, 
where 
${\lambda _{{v_i}}},{\lambda _{{\varphi _i}}},{\lambda _{{p_i}}} \in \left[ { - 1,1} \right]$.
In the NN, the \textit{Tanh} activation function is used to output ${\lambda _{{v_i}}}$, ${\lambda _{{\varphi _i}}}$ and ${\lambda _{{p_i}}}$.
In the state space of ${{{{\rm{D}}_m}}}$, $\widehat {\bf{Q}}\left[ n \right]$ and $\widehat {\bf{R}}_m^{{{\rm{cum}}}}\left[ n \right]$ have respectively $\left( {2M + 4} \right)$ and $\left| {{{\bf{C}}_m}} \right|$ dimensions, 
while 
${\left\| {{{\bf{Q}}_m}\left[ n \right] - {\bf{q}}_m^{\rm{F}}} \right\|}$,
${{v_m}\left[ n \right]}$ 
and 
${{E_m}\left[ n \right]}$ each have only one dimension.
To solve the dimension imbalance problem, a pre-diffusion network is built to extend the dimensionality of the input states for comparison with other state dimensions.
More specifically, ${\left\| {{{\bf{Q}}_m}\left[ n \right] - {\bf{q}}_m^{\rm{F}}} \right\|}$, ${{v_m}\left[ n \right]}$, and ${{E_m}\left[ n \right]}$ are extended into the ${N_{ \rm{d}}}$, ${N_{\rm{v}}}$ and ${N_{\rm{e}}}$ dimensions.
These dimensions are then combined with $\widehat {\bf{Q}}\left[ n \right]$ and $\widehat {\bf{R}}_m^{{{\rm{cum}}}}\left[ n \right]$ as inputs to the actor and critic networks \cite{WangY2022IOT, DingR2020TWC, ZhangZ2023IOT}.

\textit{\textbf{2. Critic Network}}: 
The inputs to the critic network are the extended states of all the agents and the joint actions.
The critic network goes through the global states and actions to evaluate the goodness of the agents' actions.
In the actor network, the inputs are partially observable states.
The centralized training and distributed execution training method can significantly improve the agent's learning efficiency.
The critic network has the same hidden layer structure as the actor network.
The output layer of the critic network does not utilize any activation function but directly outputs the raw values \cite{DingR2020TWC, ZhangZ2023IOT}.

\textbf{\textit{3. Target network}}: 
In the MADDPG algorithm, the target network computes the loss of the actor and critic networks to stabilize their updates.
The NN architecture of the target network is the same as the actor network. 
The critic network and the target network use soft updates to update the network parameters that can make the parameters of the target network follow the changes of the current network more smoothly, which helps to reduce the instability and oscillation in the training process and improve the stability of training.

\subsection{Training Algorithm}

\begin{algorithm}[tb]
	\caption{MADDPG-Based SCTPD Algorithm for Problem ($\mathcal{P}_{1}$)}
	\label{alg:algorithm1}
	Initialize the network, including actor network $\pi \left( {s;\theta _i^\pi } \right)$ for each agent, the target actor network with weights $\theta _i^{\pi '} = \theta _i^\pi$, the critic network $Q\left( {s,a;\theta _i^Q} \right)$ and the target critic network with weights $\theta _i^{Q'} = \theta _i^Q$.\\
	Initialize the experience replay buffer $\mathcal{C}$, mini-batch size with ${M_b}$.\\	
	\For{episodes $=0, 1,\cdots ,{M_{{\rm{ep}}}}$}{
		Initialize the starting position, speed, and initial energy consumption of multiple UAVs.\\
		\While{${E_i}\left[ n \right] > 0$}{ 
			  \For{All the agents}{
			  	Each agent inputs the partially observed state ${o_i}\left[ n \right]$ into the respective actor network,\\
			  	Get action ${a_i}\left[ n \right] = \pi \left( {{o_i}\left[ n \right];\theta _i^{\rm{\pi }}} \right) + \mathcal{N}$, where $\mathcal{N}$ is the exploration noise.			  	
			  }
		    Multiple UAVs performing joint maneuvers $a\left[ n \right]$,\\
			Update the state $s\left[ {n + 1} \right]$ and obtain the reward $r\left[ n \right]$,\\
			Store $\left\{ {s\left[ n \right],a\left[ n \right],r\left[ n \right],s\left[ {n + 1} \right]} \right\}$ in the experience replay buffer $\mathcal{C}$ \\
			\If{Memory counters $ > M$}
			{From the experience playback buffer ${\mathcal{C}}$, remove the oldest sample.}
			\For{All the agents}{
				Randomly extract mini-batch size of ${M_b}$ transitions from $\mathcal{C}$,\\
				Update the weights $\theta _i^Q$ for the critic network and $\theta _i^\pi$ for the actor network.\\
				Update the weights $\theta _i^{Q'}$ of the target critic network and the weights $\theta _i^{\pi '}$ of the target actor network.
			}
		}
	}
\end{algorithm}

The proposed SCTPD algorithm in this work is mainly divided into two parts: the training phase and the implementation phase.
In the training phase, each agent starts from its starting position and ends until it either runs out of batteries or reaches its end position, and the maximum number of training rounds is ${M_{{\rm{ep}}}}$.
Before the training starts, the respective NN parameters of all the agents are initialized.
At the beginning of each round, the starting position, speed, and energy consumption of the multi-drone are first initialized, where the initial energy of the multi-drone battery is ${E_{\max }}$.
In the $n$th time slot of each round, each agent obtains the joint action by feeding its respective partially observed state into its respective actor policy network. 
A normally distributed noise $\mathcal{N}$ with mean 0 and standard deviation 0.1 is utilized to prevent from falling into a locally optimal policy.
All agents perform this joint action to obtain a new state $s\left[ {n + 1} \right]$ and a reward $r\left[ n \right]$ and the transition tuple $\left\{ {s\left[ n \right],a\left[ n \right],r\left[ n \right],s\left[ {n + 1} \right]} \right\}$ at this point is stored in the experience playback buffer ${\mathcal{C}}$.
If the number of samples exceeds the storage space of the experience playback buffer, the oldest sample is replaced using the newest sample. 
In network updating, each agent randomly samples a fixed batch of experience from the experience playback buffer ${\mathcal{C}}$, obtains the loss of the actor network and the critic network, and updates the network parameters by a back-propagation algorithm.
The target network is then slowly updated by a soft update technique.
In the implementation phase, each agent inputs the current partial observation state into their respective actor policy networks to obtain the trajectories and transmit power.

\begin{figure*}[t]
	\centering
	\subfigure[Average cumulative reward.]{
		\label{fig03a}
		\includegraphics[width = 0.32 \textwidth]{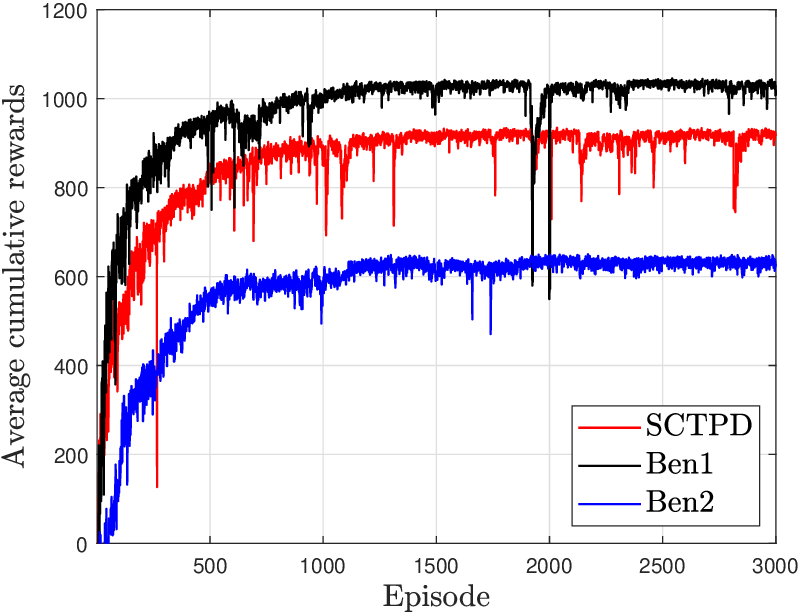}}
	\subfigure[Average cumulative security throughput.]{
		\label{fig03b}
		\includegraphics[width = 0.32 \textwidth]{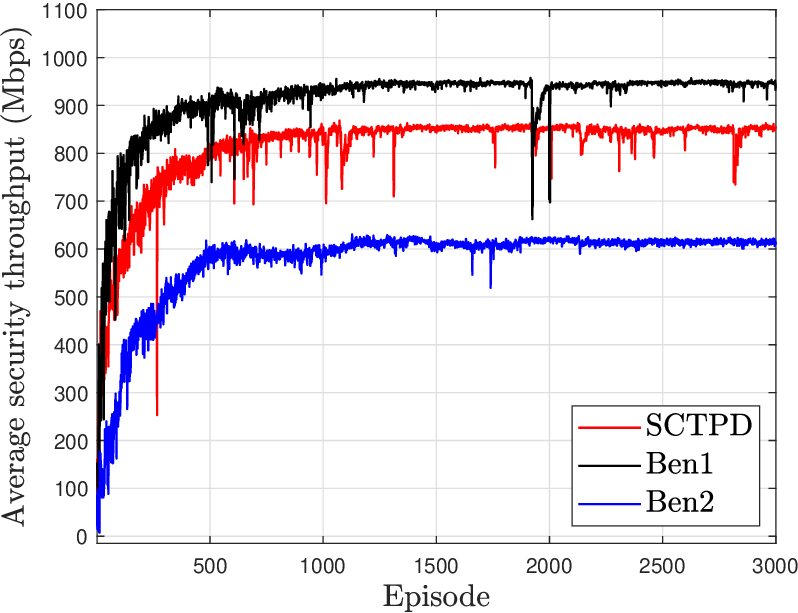}}
	\subfigure[Average fairness index.]{
		\label{fig03c}
		\includegraphics[width = 0.32 \textwidth]{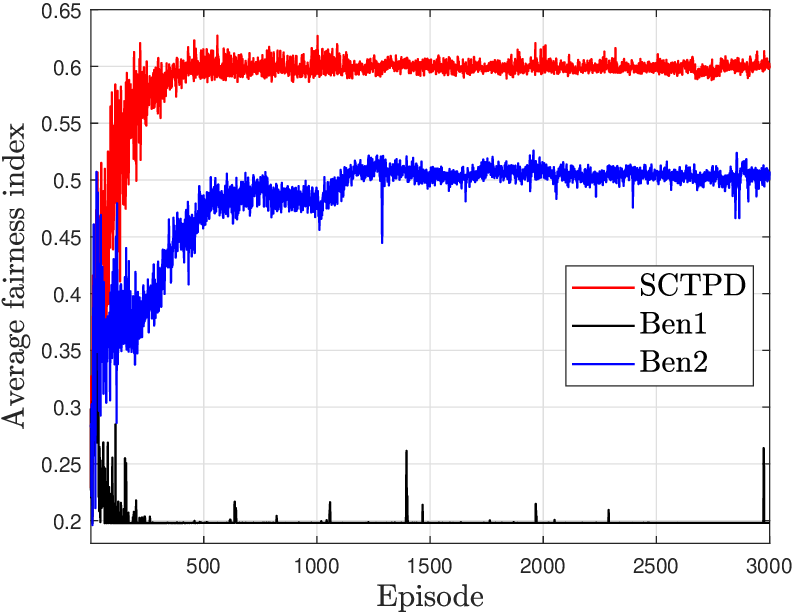}}
	\caption{The training process.}
	\label{fig03}
\end{figure*}

\begin{figure}[t]
	\centering
	\includegraphics[width = 0.4 \textwidth]{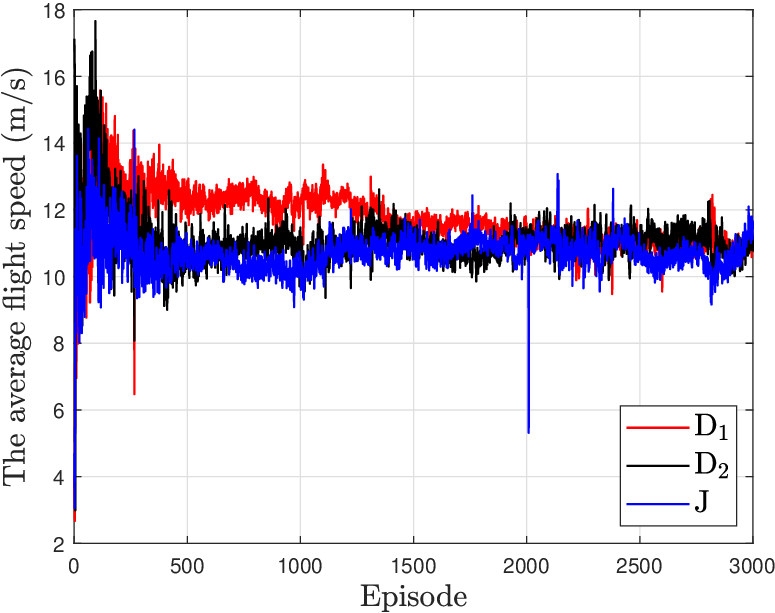}
	\caption{Average speed per episode with training.}
	\label{fig04}
\end{figure}

\section{Simulation Analysis}
\label{sec:Simulation}

\begin{table}[t]
	\caption{\emph{List of Simulation Parameters.}}
	{
		\begin{center}
			\begin{tabular}{c| c  }
				\hline
				\textbf{Notation}               & \textbf{Value}\\
				\hline
				$\left[ {{\bf{q}}_{\rm{E}}^{\rm{F}};{\bf{q}}_{\rm{E}}^{\rm{F}}} \right]$    & ${\left[ {0,300;510,300} \right]^T}$\\
				\hline
				${\bf{q}}_{{\rm D}_{1}}^{\rm{I}} = {\bf{q}}_{{\rm D}_{1}}^{\rm{F}}$    & ${\left[ {200, 0} \right]^{\rm{T}}}$\\
				\hline
				${\bf{q}}_{{\rm D}_{2}}^{\rm{I}} = {\bf{q}}_{{\rm D}_{2}}^{\rm{F}}$    & ${\left[ {300, 0} \right]^{\rm{T}}}$\\
				\hline
				${\bf{q}}_{{\rm J}}^{\rm{I}} = {\bf{q}}_{{\rm J}}^{\rm{F}}$    & ${\left[ {250, 250} \right]^{\rm{T}}}$\\
				\hline
				${\rm{H}}$                      & 70 m\\
				\hline
				${R_{\max }}$          			& 150 \\
				\hline
				${E_{\max }}$          			& 13000 J \\
				\hline
				${v_{\max }}$  		   			& 20 m/s\\
				\hline
				${{{a}}_{\max }}$       		& 5 m/s\\
				\hline
				${{{a}}_{\min }}$       		&-5 m/s\\
				\hline
				$P_{\rm{u}}^{\max}$        		&1 W\\
				\hline
				$P_{\rm{J}}^{\max}$       		&0.3 W\\
				\hline
                $D$                      		& 50 m\\
				\hline
				$B$                      		& 1 MHz\\
				\hline
				${N_0}$							& -170 dB\\
				\hline
				${\beta _0}$					& - 50 dB\\
				\hline
				${k_f}$					        & 0.95\\
				\hline
				${\eta _a},{\eta _b}$			& 12.08, 0.11\\
				\hline
				${\eta _{LoS}},{\eta _{NLoS}}$	& 1.6 dB, 23 dB\\
				\hline
				${\delta _t}$					& 1 s\\ 	
				\hline
				${\kappa _{{\mathrm{ec}}}}$					& 1 s\\ 
				\hline
				${\kappa _{{\rm{ar}}}}$					& 1 s\\ 
				\hline
				${\kappa _{{\rm{nar}}}}$					& 1 s\\ 
				\hline
				${\kappa _{{\rm{th}}}}$					& 25\\ 
				\hline
				${\kappa _{{\rm{nth}}}}$					& 1\\ 
				\hline
				${\kappa _{\rm{A}}}$ 						& -5\\ 
				\hline
				${\kappa _{\rm{L}}}$					& -10\\ 			
				\hline							
			\end{tabular}
		\end{center}
	}
	\label{tab2}
\end{table}

In this subsection, the simulation results with $M =  2$ and $K =  10$ are presented to verify the feasibility of the proposed SCTPD algorithm and evaluate its performance.
The detailed information on the parameter settings of the multi-UAV communication system is shown in Table \ref{tab2}. 
%The simulation parameters related to NNs are specified as follows. 
In the SCTPD algorithm, ${{{\rm{D}}_m}}$ and ${\rm{J}}$ are considered agents, and each agent actor network possesses four NNs. 
The actor network structure for each agent consists of three hidden layers, each with 256, 128, and 64 neurons, respectively.
The extended dimensions of which states are ${N_{\rm{d}}} = 2$, ${N_{\rm{v}}} = 2$ and ${N_{\rm{e}}} = 2$, respectively.
The actor network for each agent outputs the velocity vector and transmit power in each time slot, so the output layer has three neurons.
To prevent the gradient from vanishing, the hidden layers of all the agents use the \textit{Relu} function as the activation function.
The hidden layer setup of the critic network is the same as that of the actor network, with the difference that the input of the critic network consists of the current states and joint actions of all the agents, and the output dimension is 1.
Both the actor network and the critic network are feed-forward fully connected NNs, and the networks are updated using the ADAM optimizer \cite{KPD2013ICLR} with a learning rate of 0.001 and a discount factor $\gamma$ of 0.9.
The size of the empirical replay buffer ${\mathcal{C}}$ is $4 \times {10^4}$, and the size of the trained small batch data ${M_{\rm{b}}}$ is 512.

To verify the effectiveness of the proposed SCTPD algorithm, two schemes are used as benchmarks and the performance is compared with that of the proposed algorithm, which are described as follows.

\begin{enumerate}	
	
	\item Benchmark I (`Ben1'): In this scheme, the communication UAVs only consider maximizing the total secrecy throughput $\left( {R^{{\rm{Ben1}}}} \right)$, defined as
	\begin{align}
			{R^{{\rm{Ben1}}}} = {\sum\limits_{n = 1}^N {\sum\limits_{m = 1}^M {\sum\limits_{{u_i} \in {{\rm{C}}_m}} {b_{m,{u_i}}}\left[ n \right]R_{m,{u_i}}^{\sec }\left[ n \right]{\delta _t}} } }.
			\label{totalsecthroughput}
	\end{align}
	 Specifically, UAV will communicate with the user with the maximum security rate with the help of ${\rm{J}}$ and utilize the same hyperparameter and reward design as the SCTPD algorithm.	
	
	\item Benchmark II (`Ben2'): In this scheme, without the help of ${\rm{J}}$,  the trajectory and power of ${{{\rm{D}}_m}}$ are jointly optimized to maximize the total FST $\left( {{R^{{\rm{sum}}}}} \right)$ of the system. The same hyperparameter and reward design as the SCTPD algorithm is used.
	
\end{enumerate}

Fig. \ref{fig03} shows the trends of the average cumulative reward $\left( {r_{{\rm{average}}}^{{\rm{cum}}}} \right)$, 
the average cumulative secrecy throughput $\left( {R_{{\rm{average}}}^{{\rm{cum}}}} \right)$,
and the average fairness index $\left( {{f_{{\rm{average}}}}} \right)$, 
which are defined as
\begin{align}
		r_{{\rm{average}}}^{{\rm{cum}}} &= \frac{1}{{M + 1}}\sum\limits_{n = 1}^N {\left( {{r_{\rm{J}}}\left[ n \right] + \sum\limits_{m = 1}^M {{r_m}\left[ n \right]} } \right)}, \\
		R_{{\rm{average}}}^{{\rm{cum}}} &= \frac{1}{M}\sum\limits_{m = 1}^M {R_m^{{\rm{cum}}}\left[ N \right]}, \\
		{f_{{\rm{average}}}} &= \frac{1}{{MN}}\sum\limits_{m = 1}^M {\sum\limits_{n = 1}^N {{{\hat f}_m}\left[ n \right]} }. 	\label{avefairnessindex}
\end{align}
It can be observed that the reward curve converges for all schemes as the number of training rounds increases. 
But because of the exploration noise, the reward fluctuates in a range. 
From Figs. \ref{fig03a} and \ref{fig03b}, one can find that the average cumulative secrecy throughput of SCTPD exceeds that of Ben2 but is lower than that of Ben1. 
The reason is that fairness between users is not considered in Ben1, so communication UAVs do not need to fly frequently between users but tend to continue communicating with a user with maximum performance.
Therefore, Ben1 achieves greater security throughput and cumulative reward than the SCTPD scheme, but this comes at the expense of user fairness. 
Based on Fig. \ref{fig03c}, it can be observed that the average fairness index of the SCTPD scheme has a significant improvement compared to Ben1, which is due to the fact that under the fair scheduling scheme, the communication UAVs tend to visit all the users in the cluster, thus going to ensure fairness among the users. Specifically, to ensure fairness among users, part of the secrecy throughput is lost such that there is a trade-off between fairness and secrecy throughput.
In Ben2, aerial eavesdropping capability cannot be suppressed by the jamming UAV, which results in a smaller cumulative secrecy throughput of the communication UAV during the flight cycle. 
However, it is worth noting that Ben2's average fairness index significantly improves compared to Ben1 because a fair scheduling scheme is considered in Ben2.

Fig. \ref{fig04} plots the trend of the {average} speed of UAVs as the number of training rounds increases. 
It can be observed that as the number of training rounds increases, the average speed first increases and then decreases and finally converges to about 11 m/s. 
This is because, under the energy consumption model, the UAV consumes the least energy per slot when it flies at a speed of about 10 m/s so that it can fly for a longer time and maximize the cumulative secrecy throughput.

\begin{figure*}[t]
	\centering
	\subfigure[The actor loss of each agent.]{
		\label{fig05a}
		\includegraphics[width = 0.4 \textwidth]{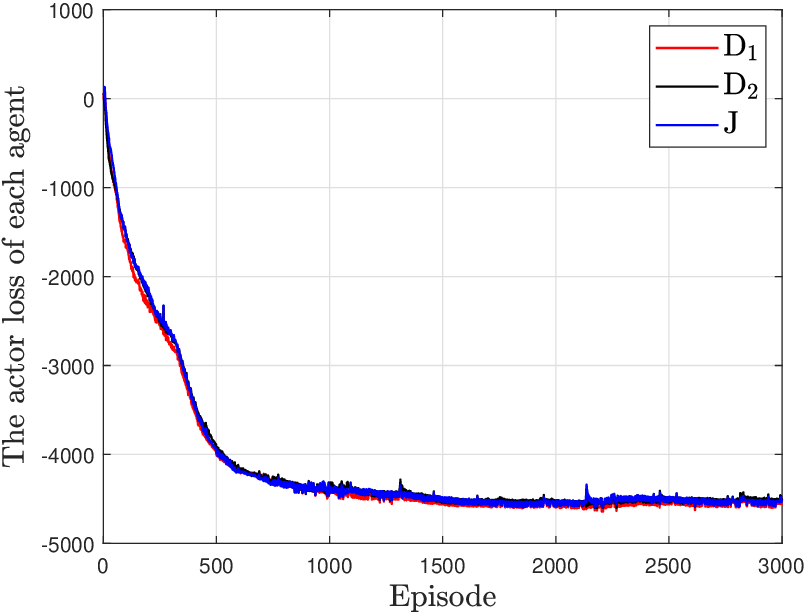}}
	\subfigure[The critic loss of each agent.]{
		\label{fig05b}
		\includegraphics[width = 0.4 \textwidth]{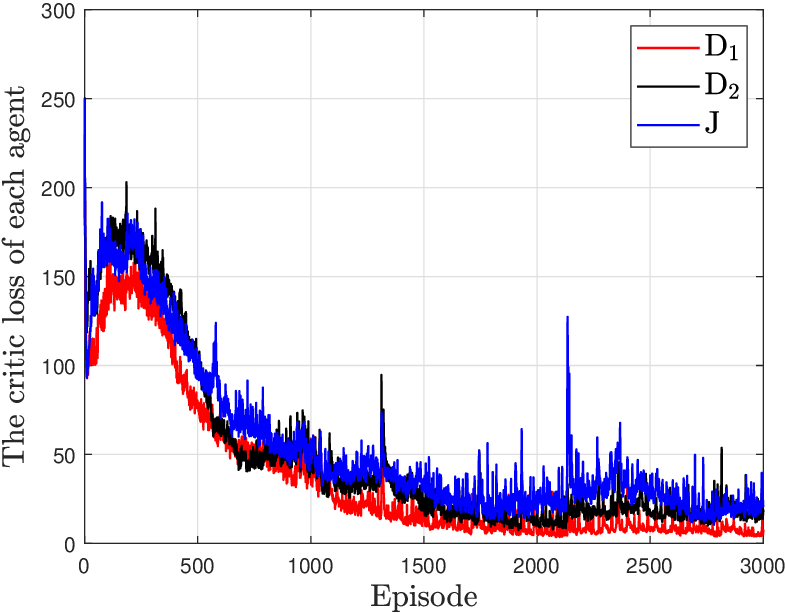}}
	\caption{Training loss per episode.}
	\label{fig05}
\end{figure*}
\begin{figure*}[t]
	\centering
	\subfigure[The probability of acceleration constraint violation.]{
		\label{fig06a}
		\includegraphics[width = 0.4 \textwidth]{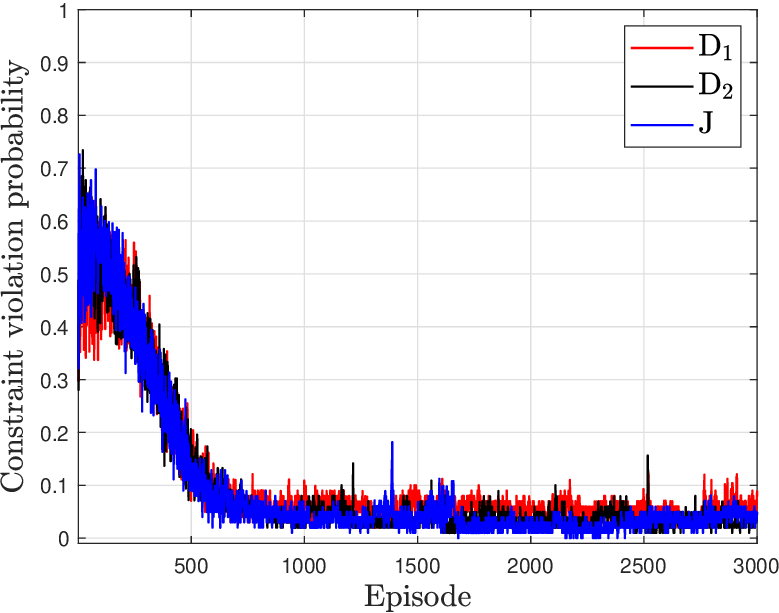}}
	\subfigure[The probability of the distance constraint violation.]{
		\label{fig06b}
		\includegraphics[width = 0.4 \textwidth]{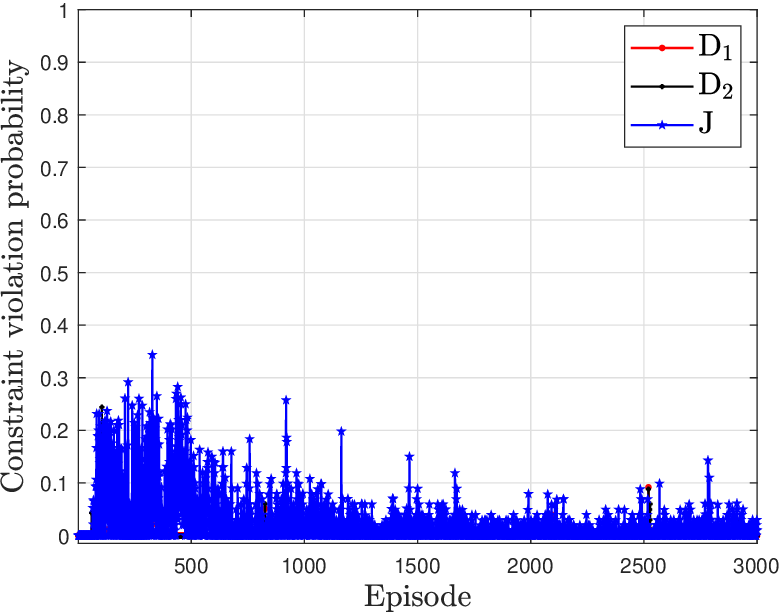}}
	\caption{The probability of constraint violation.}
	\label{fig06}
\end{figure*}

Figs. \ref{fig05a} and \ref{fig05b} show the trend of cumulative losses of the actor network and critic network of communication and jamming UAVs as the number of training rounds increases, respectively. 
It can be observed that as the number of training rounds increases, the cumulative loss of both the actor network and the critic network gradually decreases and then converges to a constant. 
The cumulative loss of the critic network converges to zero, while that of the actor network converges to a negative number.

Fig. \ref{fig06} shows the curves of constraint violation probability of the communication and jamming UAVs as the number of training rounds increases.
It can be observed from Fig. \ref{fig06a} that, at the beginning of training, the probability of violating the acceleration constraint is around 0.6 to 0.4. 
As the number of training rounds increases, the probability of violating the acceleration constraint converges at about 700 rounds and converges between 0 and 0.1. 
This is because multiple UAVs will receive a discrete punitive reward if they violate the constraint in each flight slot. 
So that the strategy network of each agent learns how to output the appropriate flight speed and angle to meet the constraints, thereby avoiding punishment.
It can be observed in Fig. \ref{fig06b} that the probability of communication UAV for violation distance constraint tends to be 0. 
However, the probability of J for violation distance constraints first increases and then decreases, and finally converges.  
This is because the user clustering algorithm is used in this work to divide users into $M$ clusters, which are then associated with multiple communication UAVs so that the distance constraint between communication UAVs is satisfied. 
To fight with the aerial eavesdropper, $J$ tends to get close to $E$. 
Therefore, as the number of training rounds increases, the probability of violation distance constraint rises first. 
Then, due to the punitive reward, the probability of violation distance constraint decreases and finally converges.
It should be noted that there is a trade-off between constraint rewards and other rewards. When the weight of the constraint reward is too large, it may affect the secrecy communication and the learning about the return to the endpoint task.

\begin{figure*}[t]
	\centering
	\subfigure[The velocity of ${{{\rm{D}}_1}}$.]{
		\label{fig07a}
		\includegraphics[width = 0.32 \textwidth]{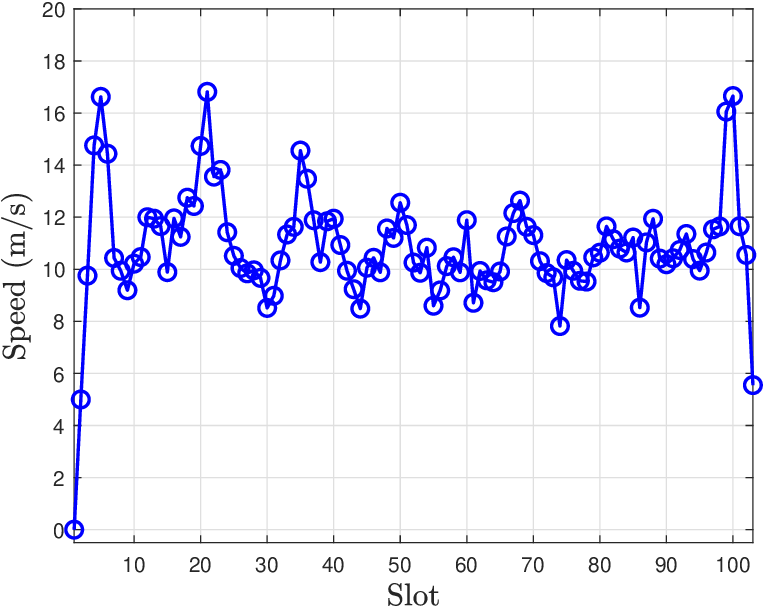}}
	\subfigure[The velocity of  ${{{\rm{D}}_2}}$.]{
		\label{fig07b}
		\includegraphics[width = 0.32 \textwidth]{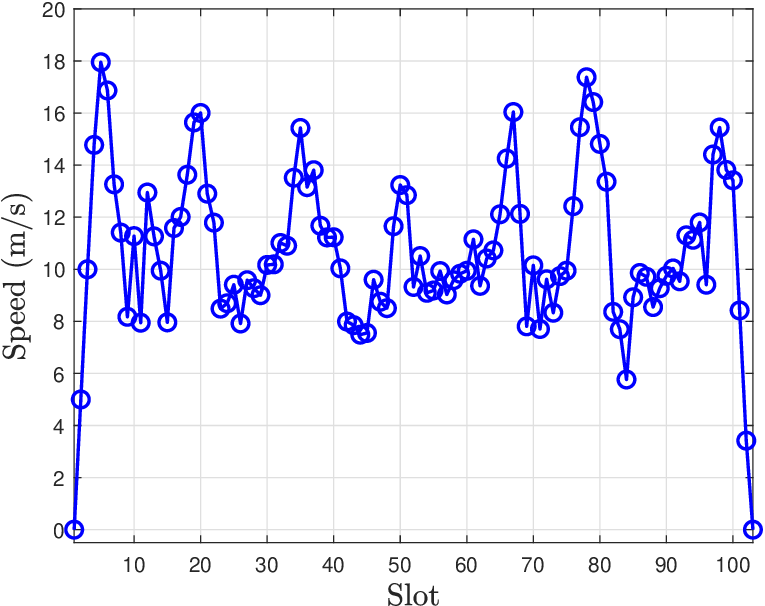}}
	\subfigure[The velocity of  ${{{\rm{J}}}}$.]{
		\label{fig07c}
		\includegraphics[width = 0.32 \textwidth]{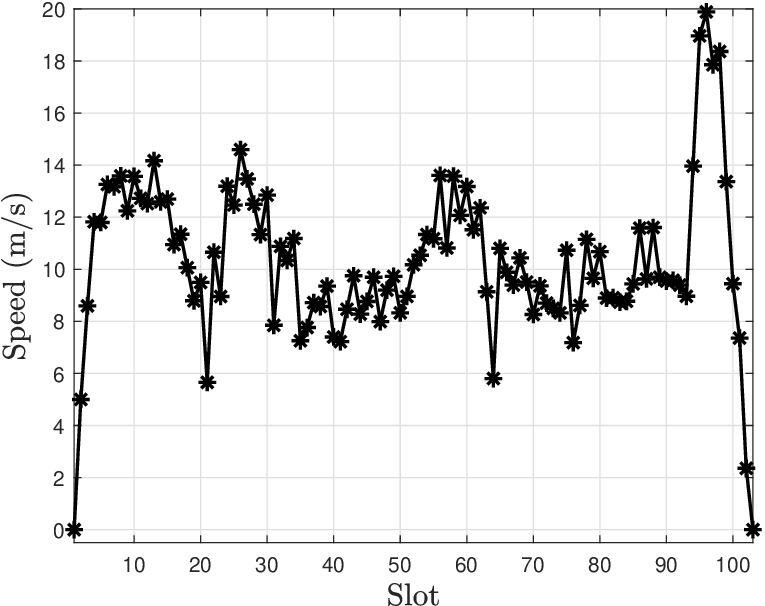}}
	\caption{The speed during the flight cycle.}
	\label{fig07}
\end{figure*}

Fig. \ref{fig07} plots the speed of communication and jamming UAVs during flight.
From Fig. \ref{fig07a} and Fig. \ref{fig07b}, it can be observed that the flight speed of the communication UAV tends to fly at a speed of 16 m/s far from the users, while it flies at a speed of about 10 m/s near the users. 
This is because, under the energy consumption model, the UAV has the lowest energy consumption per time slot at 10 m/s and the longest flight distance when it tends to 18 m/s \cite{ZengY2019TWC}. 
Therefore, by flying at the speed shown in Fig. \ref{fig07}, it is possible to fly as long as possible and to fly over the user as quickly as possible to maximize the cumulative secrecy throughput. 
It can be observed from Fig. \ref{fig07c} that the speed of the jamming UAV first increases and then tends to fly at a speed of about 10 m/s. 
This is because, first, J needs to fly as quickly as possible to the vicinity of $E$ to reduce the SINR of E. 
Then, it flew around $E$ at a speed of about 10 m/s.

\begin{figure*}[t]
	\centering
	\subfigure[User scheduling for $D_1$.]{
		\label{fig08a}
		\includegraphics[width = 0.235 \textwidth]{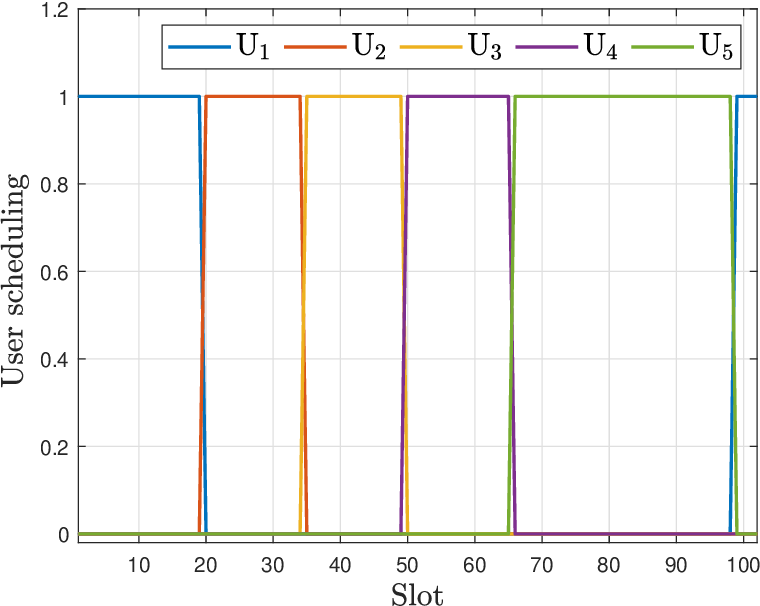}}
	\subfigure[User scheduling for $D_2$.]{
		\label{fig08b}
		\includegraphics[width = 0.235 \textwidth]{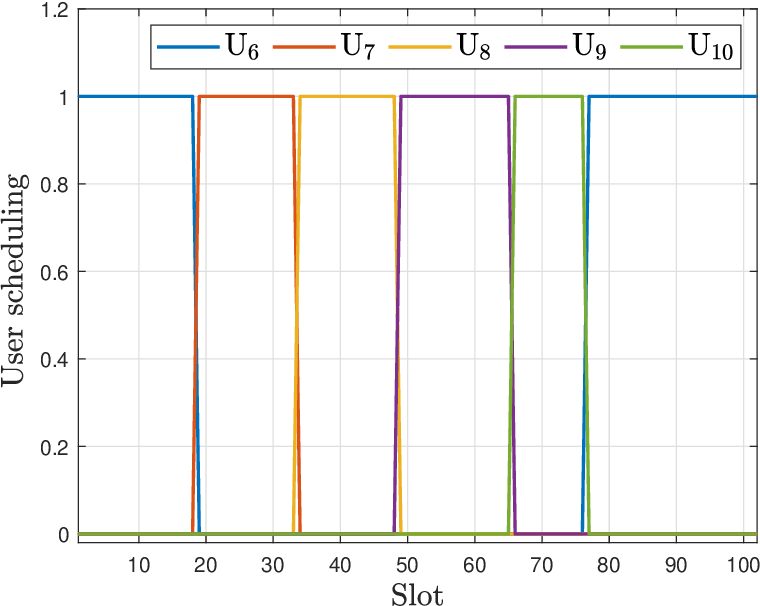}}
	\subfigure[Secrecy rate for $D_1$.]{
		\label{fig08c}
		\includegraphics[width = 0.235 \textwidth]{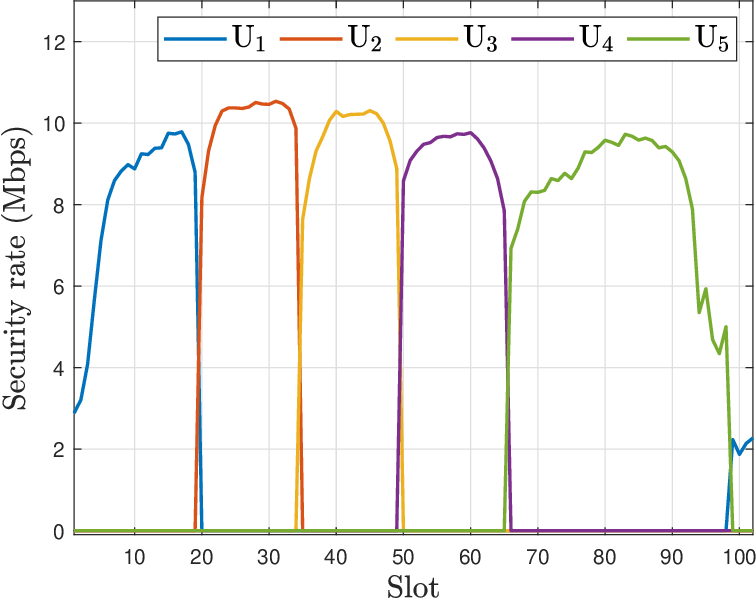}}
	\subfigure[Secrecy rate for $D_2$.]{
		\label{fig08d}
		\includegraphics[width = 0.235 \textwidth]{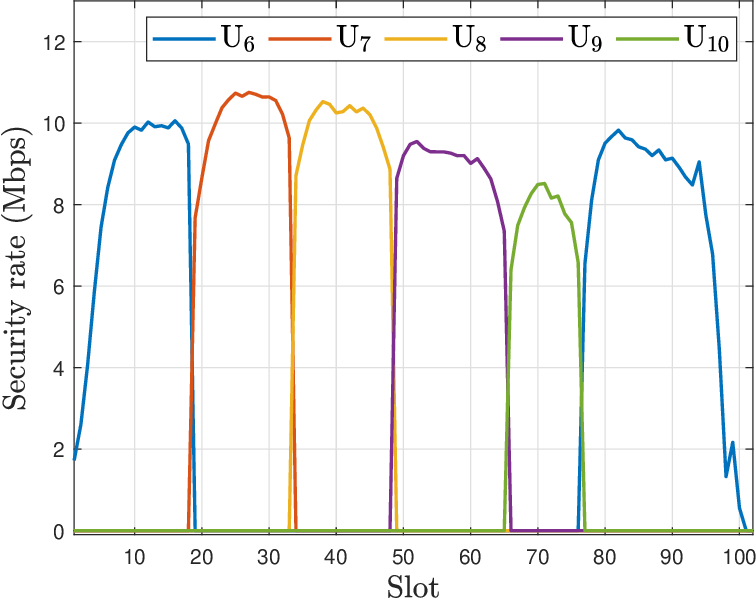}}
	\caption{User scheduling and secrecy rate during the flight cycle.}
	\label{fig08}
\end{figure*}

Fig. \ref{fig08} shows the user scheduling and secrecy rates of communication UAV during the flight cycle.
It can be observed from Figs. \ref{fig08a} and \ref{fig08b} that all users in the cluster associated with the communication UAV are awakened in turn to schedule the users with the highest fair secrecy throughput in each time slot so as to ensure fairness among users.
Based on Figs. \ref{fig08c} and \ref{fig08d}, we find that the secrecy rate of the communication UAV in each time slot always increases first and then decreases because the communication UAV approaches and moves away the user.

\begin{figure*}[t]	
	\centering	
	\subfigure[The flight trajectory of the UAVs.] {
		\label{fig10a}
		\includegraphics[width = 0.32 \textwidth]{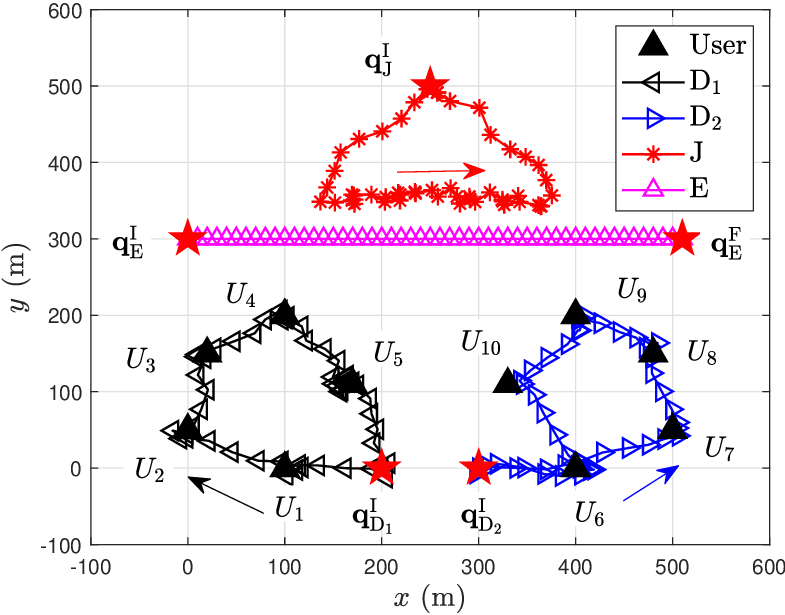}}
	\subfigure[The power of the UAVs.] {
		\label{fig10b}
		\includegraphics[width = 0.32 \textwidth]{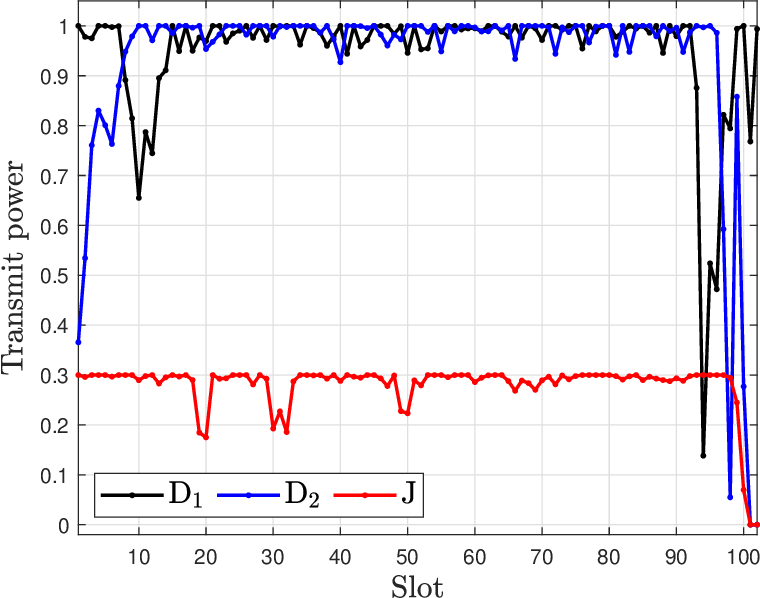}}
	\subfigure[The average FST.] {
		\label{fig10c}
		\includegraphics[width = 0.32 \textwidth]{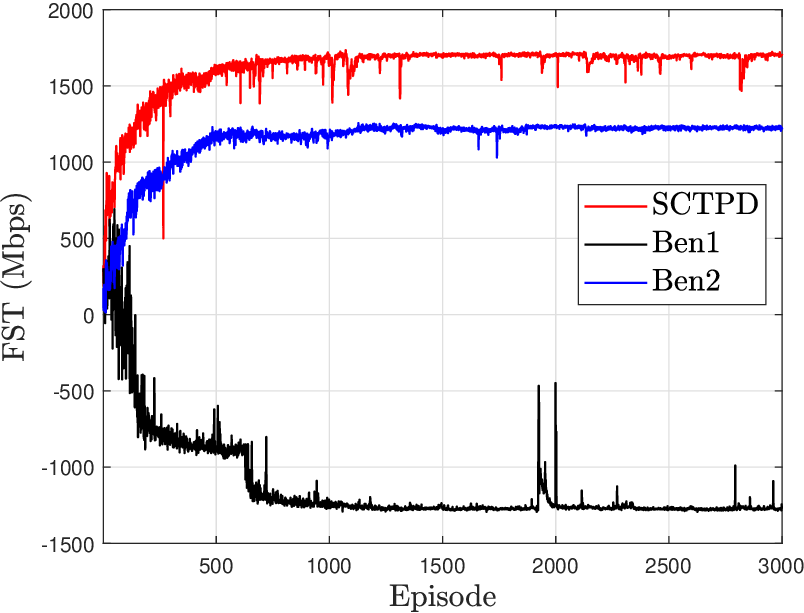}}
	\caption{The optimal trajectory, transmit power, and FST. }
	\label{fig10}	
\end{figure*}

Fig. \ref{fig10a} shows the two-dimensional trajectories of communication and jamming UAVs. 
Upon observation, it becomes apparent that the ground users are segregated into two clusters, with each cluster being served by a dedicated communication UAV.
Each communication UAV starts from the starting point (marked with $\star$), flies to the user's surroundings in turn, and tries to fly as far away from E as possible, providing secrecy and fair services for the ground users.
When the energy is about to be exhausted, it returns to the destination. 
J flies from the starting point to send jamming signals around E to suppress the eavesdropping ability and tends to follow E's flight trajectory to fly and maintain a certain safe distance. 
Finally, if energy consumption is about to run out, J will return to the endpoint. 
Fig. \ref{fig10b} shows the transmit power of all the UAVs. 
The starting point of $ D_1 $ and $ D_2 $ is very close, considering the mutual interference, so $ D_1 $ and $ D_2 $ work with a lower power. 
When flown to a relatively long distance, they transmit signals at approximately the maximum power. 
Moreover, it can be clearly found that near $U_{10}$, to reduce the interference to $U_5$ served by $D_1$, $D_2$  is more inclined to serve $U_6$. 
At the beginning, $J$ transmits the AN with the maximum power. When near the scheduled users, $J$ works with the lower power to reduce interference. 
An interesting finding is that, rather than adjusting the transmit power, UAVs are more inclined to adjust trajectories to improve communication performance.
Fig. \ref{fig10c} demonstrates the FST for all schemes as the number of training rounds increases. 
It can be observed that the FST curve converges for all schemes as the number of training rounds increases, and the SCTPD scheme has a greater average cumulative FST than the base schemes. 
In Ben1, since user fairness is not considered, its FST is the worst.

\begin{figure}[t]
	\centering
	\includegraphics[width = 0.4 \textwidth]{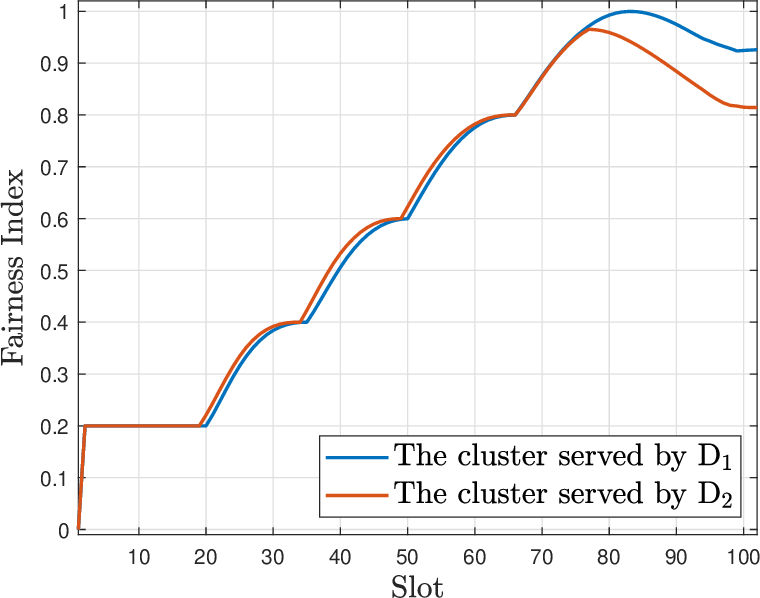}
	\caption{Fairness index of the clusters.}
	\label{fig09}
\end{figure}

Fig. \ref{fig09} shows the fairness index of communication UAVs during flight. 
One can see that the fairness index goes up and then goes down to a constant. 
This is because communication UAVs need to visit each user first to ensure fairness constraints. 
When the communication UAV provides services to more users, the fairness index of the considered system increases with the increase in the number of users in the serviced cluster, reflecting the fairness of the communication service. 
When fairness constraints are met, communication UAV will continue to communicate around certain users to maximize the total secrecy rate, resulting in a decline in the fairness index.

\section{Conclusion}
\label{sec:Conclusions}

In this work, the security problem of multi-UAV-aided communication systems was studied in the aerial eavesdropping scenario. 
To improve the security of the considered system, a friendly UAV was utilized to transmit AN signals to suppress the aerial eavesdropper. 
Considering the limited energy consumption and flight conditions of UAVs and the fairness among the users, we formulated an optimization problem. 
Firstly, all the users were clustered based on the $K$-means algorithm.  
Then, a new algorithm, named SCTPD, was proposed to maximize the total FST by jointly optimizing the trajectory and transmission power of all the UAVs. 
In addition, adaptive energy thresholds were taken into account in the reward design to balance return to the endpoint and secure communication. Simulation results verified the algorithm's convergence and effectiveness.

%3D

%Multi-antenna technology and agent reflector have great potential to improve the performance of communication networks. 
%Therefore, 
%In future work, how to use multi-agent reinforcement learning to solve the problem of multi-UAV trajectory design and resource allocation under multi-antenna technology is an interesting problem. 
%The goal is to provide users with higher quality and more secure communication services to meet the changing challenges in the DRone-assisted communication environment.

\end{document}